\documentclass[prb,nobalancelastpage,twocolumn,nofootinbib,superscriptaddress,showpacs]{revtex4-1}

\usepackage{color,amsthm,amsmath,amsfonts,graphicx,bm}
\usepackage{bm,eufrak}
\usepackage{amsfonts}
\usepackage{amssymb}
\usepackage{amsmath,mathrsfs}
\usepackage{epsfig}
\usepackage{graphicx}
\usepackage{enumerate}
\usepackage{multirow}
\usepackage{verbatim}
\usepackage{subfigure}
\usepackage{bbold}
\usepackage{soul}
\usepackage{scrextend}
\usepackage{tikz}
\usetikzlibrary{decorations.pathreplacing}

\newcommand{\spec}{\mathrm{spec}}
\newcommand{\mc}[1]{\mathcal{#1}}
\newcommand{\mr}[1]{\mathrm{#1}}
\newcommand{\dg}{\dagger}

\newcommand{\mb}{\mathbf}

\newcounter{notes}
\stepcounter{notes}

\DeclareFontFamily{OT1}{pzc}{}
\DeclareFontShape{OT1}{pzc}{m}{it}{<-> s * [1.10] pzcmi7t}{}
\DeclareMathAlphabet{\mathpzc}{OT1}{pzc}{m}{it}

\begin{document}

\title{Tensor networks demonstrate the robustness of localization and symmetry protected topological phases}

\date{\today}

\author{Thorsten B. Wahl}
\affiliation{Rudolf Peierls Centre for Theoretical Physics, Clarendon Laboratory, Parks Road, Oxford OX1 3PU, United Kingdom}




\begin{abstract}
We prove that all eigenstates of many-body localized symmetry protected topological systems with time reversal symmetry have  four-fold degenerate entanglement spectra in the thermodynamic limit. To that end, we employ unitary quantum circuits where the number of sites the gates act on grows linearly with the system size. We find that the corresponding matrix product operator representation has similar local  symmetries as matrix product ground states of symmetry protected topological phases. Those local  symmetries give rise to a $\mathbb{Z}_2$ topological index, which is robust against arbitrary perturbations so long as they do not break time reversal symmetry or drive the system out of the fully many-body localized phase.

\end{abstract}

\maketitle

\section{Introduction} 
The idea that systems out of equilibrium act as their own heat bath was first challenged by Anderson in 1958~\cite{anderson1958absence}. Later works confirmed rigorously that in non-interacting one and two dimensional systems (without broken time reversal symmetry or spin orbit coupling) arbitrarily weak disorder leads to localization of all single particle eigenstates~\cite{Lee_RMP1985}.
Strikingly, in one dimension, the resulting lack of transport survives for sufficiently strong disorder if interactions are included~\cite{Fleishman1980,gornyi2005interacting,basko2006metal,imbrie2016many}. 
 Such many-body localized (MBL) systems~\cite{NandkishoreHuse_review,AltmanReview,Luitz2017,ImbrieLIOMreview2017,Abanin2017,Alet2017} retain a memory of their initial state for arbitrarily long times, thus violating the eigenstate thermalization hypothesis (ETH)~\cite{1984Peres,deutsch1991quantum,srednicki1994chaos,srednicki1999,Rigol:2008bh,DAlessio2015aa,Borgonovi2016}. Many-body localization was observed in recent optical lattice experiments on one~\cite{Schreiber842} and two dimensional systems~\cite{Choi1547,bordia2017quasiperiodic2D}. Numerical studies predict other exotic phenomena in MBL systems, such as the logarithmic growth of entanglement following a quantum quench~\cite{Chiara2006,Bardason2012,serbyn2013universal,Luitz2016,Singh2016,Zhou2017} and an unconventional transition to the thermal phase~\cite{oganesyan2007localization,Berkelbach2010,pal2010mb,Pekker2014HG,Vosk2014,VHA_MBLTransition,Potter:2015ab,Luitz2015,Devakul2015,Gopalakrishnan2016,nidari2016,AltmanReview,SondhiQPT_review,yu2016bimodal,Zhang2016,Agarwal2017RareregionReview,Khemani2017MBLT,Parameswaran2017,Khemani2017,sachdev2007quantum}. From a conceptual point of view, MBL systems are characterized by an extensive number of local integrals of motion (LIOM)~\cite{serbyn2013local,chandran2015constructing,ros2015integrals,Inglis_PRL2016,Rademaker2016LIOM,Monthus2016,Pekker2017,ImbrieLIOMreview2017,Goihl2017,Abi2017,Geraedts2017,Monthus2017} and area-law entangled eigenstates~\cite{2013Bauer_Nayak,PalThesis,Friesdorf2015}. Excited eigenstates thus have similar features as the ground states of local gapped Hamiltonians~\cite{2006Hastings}, which is why those eigenstates can be efficiently approximated by matrix product states (MPS)~\cite{PerezGarcia2007,znidaric2008many,2013Bauer_Nayak,Khemani_MBL_MPS,Kennes2016,Lim2016MPS,Yu2017,Devakul2017}. Moreover, unitary quantum circuits (a special type of tensor networks~\cite{PerezGarcia2007,PEPS,2009Cirac_review,2013Eisert_review,2013Orus_review}) encode the entire set of eigenstates efficiently~\cite{Pekker2017MPO,Chandran2015STN,Pollmann2016TNS,Wahl2017PRX}. 

The absence of thermal fluctuations in MBL systems facilitates symmetry breaking orders and symmetry protected topological (SPT) orders at all energy scales, which in clean systems can only exist at zero temperature~\cite{2013Bauer_Nayak,Huse2013LPQO,Chandran2014SPT,kjall2014many,bahri2015localization,2015Slagle,2015Potter}. 
 Hence, in the localized case all eigenstates can be SPT, which makes MBL systems viable candidates for topological quantum memories at arbitrary energy density~\cite{2013Bauer_Nayak}. Symmetry and localization protected systems thus interface with quantum information theory both at their theoretical description by tensor networks and their practical potential for quantum information storage and processing tasks~\cite{2013Bauer_Nayak,Chandran2014SPT}.
 
One of the greatest accomplishments in tensor network research so far was the classification of all gapped topological phases in one dimension~\cite{Pollmann2010,2011Chen,2011Schuch}. This was made possible by the insight that ground states of one-dimensional gapped systems can be efficiently approximated by MPS~\cite{MPS_faithful,2015Huang}. This suggests that tensor networks might also be used to classify SPT MBL phases as proposed in Ref.~\onlinecite{Chandran2014SPT}. 

In this article, we establish tensor networks as a tool for such a classification. Specifically, we use quantum circuits to prove that MBL phases in one dimension protected by time reversal symmetry fall in two different classes given by a $\mathbb{Z}_2$ topological index. The only assumption we make in our proof is that a two-layer unitary quantum circuit~\cite{Wahl2017PRX} diagonalizes the MBL Hamiltonian exactly in the thermodynamic limit if the length of the gates increases linearly with the system size. As we argue, this applies to MBL systems as defined above, also known as fully many-body localized (FMBL) systems~\cite{huse2014phenomenology}, which do not possess a mobility edge~\cite{Luitz2015,DeRoeck2016}. We find that the global time reversal symmetry of the system gives rise to local symmetries of the tensors - similarly to MPS with a global symmetry~\cite{Sanz2009,Pollmann2010}. We also prove that the topological index determined by those local symmetries is robust against arbitrary symmetry respecting perturbations as long as they do not drive the system out of the FMBL phase. Finally, we show that all eigenstates in the SPT MBL phase have four-fold degenerate entanglement spectra. 

In the following Section we give a very brief introduction into symmetry protected topological many-body localized phases. Sec. III provides a summary of the main results and an intuitive (non-technical) outline of the stability proof, which follows in Sec. IV. Sec. V concludes the paper and gives outlook for future work. 

Those readers interested only in the general MBL classification idea using tensor networks and the physical implications may skip Sec. IV.


\section{Symmetry and localization protected phases}

\subsection{Local integrals of motion}  

Throughout this article, we consider a disordered spins chain in one dimension with periodic boundary conditions. For sufficiently strong disorder, where the system is in the FMBL phase, the Hamiltonian commutes with an extensive number of LIOMs $\tau^i_z$, $[H, \tau_z^i] = [\tau_z^i, \tau_z^j] = 0$.  $\tau_z^i$ is related to $\sigma_z^i$ (Pauli-$z$ operator at site $i$) by a quasi-local unitary transformation $U$, i.e., $\tau_z^i = U \sigma_z^i U^\dg$ is an effective spin  exponentially localized around site $i$~\cite{chandran2015constructing}. Note that $U$ also diagonalizes the Hamiltonian. 
The eigenstates can be labelled by the eigenvalues of the $\tau_z^i$ operators,  known as l-bits.  The decay length $\xi_i$ of $\tau_z^i$ depends on the specific disorder realization. 
In the FMBL phase, the likelihood of finding a decay length of order $\mc{O}(N)$ is zero in the limit $N \rightarrow \infty$~\cite{chandran2015constructing,Abi2017}. 



\subsection{Symmetry and localization protected phases} 

In FMBL systems, all eigenstates fulfill the area law of entanglement. This allows, in principle, for the topological symmetry protection of the full set of eigenstates. In one dimensional systems, time reversal symmetry or on-site symmetries given by an Abelian symmetry group~\cite{2016Potter_Vasseur} are candidates. (Note that as opposed to ground states of clean systems,  for random disordered systems, inversion symmetry is not an option.) In this article, we will show the robustness of time reversal symmetry protected MBL systems and point out what currently prevents the generalization to on-site symmetry groups (see Sec. III). 

As a paradigmatic example, consider the disordered cluster model with random couplings~\cite{bahri2015localization},
\begin{align}
H = \sum_{i=1}^N \left(\lambda_i \sigma^{i-1}_x \sigma_z^i \sigma^{i+1}_x + h_i \sigma^i_z + V_i \sigma^i_z \sigma^{i+1}_z \right)
\label{eq:top_Ham}
\end{align}
on a chain with $N$ sites and periodic boundary conditions. (We define position indices modulo $N$.)  $\lambda_i$, $h_i$ and $V_i$ are real and chosen independently from a Gaussian distribution with standard deviation $\sigma_\lambda$, $\sigma_h$ and $\sigma_V$, respectively. In Ref.~\onlinecite{bahri2015localization} it was observed numerically that the entanglement spectra of all eigenstates are approximately four-fold degenerate for $\sigma_h, \sigma_V \ll \sigma_\lambda$ and finite $N$. We prove the exact degeneracy in the limit $N \rightarrow \infty$ using the fact that the corresponding  SPT MBL phase is protected by time reversal symmetry, which in this case is a combination of complex conjugation ($^*$) and rotation by $\sigma_z$,
\begin{align}
H = \sigma_z^{\otimes N} H^* \sigma_z^{\otimes N}. \label{eq:sym}
\end{align}
In general, time reversal acts as $\mc{T} = K \mathpzc{v}^{\otimes N}$, where $K$ denotes complex conjugation and $\mathpzc{v}$ an on-site unitary operation with $\mathpzc{v} \mathpzc{v}^* = \pm \mathbb{1}$. Note that the sign will not affect the topological classification~\cite{2011Chen}, as the overall unitary $\mathpzc{v}^{\otimes N}$ fulfills $\mathpzc{v}^{\otimes N} \mathpzc{v}^{* \otimes N} = (\pm 1)^{N} \mathbb{1}$, which is $\mathbb{1}$ for even $N$. (If $N$ is odd, one can always add a completely decoupled auxiliary spin to the chain, which would not change the fact that the system is MBL.)

\vspace{12pt}
\section{Non-technical Summary of Results and Intuitive Outline of the Proof}

Numerical evidence indicates that two-layer quantum circuits with long gates approximate FMBL systems efficiently~\cite{Wahl2017PRX}. For disordered systems, they are thus the full-spectrum analogoues of matrix product states (MPS) for clean systems. This suggests that they might also be used for the classification of symmetry protected  MBL systems - as MPS were for clean systems. In this article, we provide evidence for this conclusion by using one-dimensional quantum circuits to show that MBL systems protected by time reversal symmetry fall in two different classes, where one of them is topologically non-trivial as exemplified by a four-fold degeneracy of the entanglement spectrum of \textit{all} of its eigenstates. 

The only assumption (other than being in a time reversal symmetric MBL phase) that goes into the proof is that the local integrals of motion can be represented efficiently by a quantum circuit with long gates. This is basically equivalent to not having any LIOM with a decay length of the order of the system size, i.e., to be in the FMBL phase. We show that, as a result, the Hamiltonian belongs to one of two topologically inequivalent phases. We show that it is impossible to connect the two phases adiabatically without violating either the time reversal symmetry or the FMBL condition. This is very reminiscent of SPT ground states of clean Hamiltonians: As long as the symmetry is preserved, they cannot be adiabatiacally connected to the trivial phase unless they become delocalized (having algebraically decaying correlations), i.e., the gap of the Hamiltonian closes. This is why MPS can be used for their classification: MPS always have exponentially decaying correlations and represent ground states of local gapped Hamiltonians.
If the tensors of two (symmetric) MPS cannot be continuously connected, it is impossible to connect the ground states they approximate continuously without encountering a quantum phase transition, at which correlations decay algebraically (which cannot be captured exactly by an MPS).

In the same way, the transition between the two topologically inequivalent MBL phases must lie outside the realm of systems that can be approximated efficiently by quantum circuits. Hence, at the transition, at least one LIOM must become delocalized (which does not imply the transition resembles an MBL-to-thermal transition~\cite{kjall2014many}). This correspondence between MPS classifications of ground states and quantum circuit classifications of MBL phases is summarized in Table~\ref{table}.
\begin{widetext}
\begin{center}
\begin{table}
\begin{tabular}{ |c| c| c |}
\hline
 Property & MPS & Quantum Circuit \\ 
 \hline
 & & \\
 Description of&Ground states&All eigenstates \\
 &&\\
 System&Translationally invariant\footnote{MPS can be straightforwardly defined for non-translationally invariant systems, but using them for a classification of phases in such a case requires additional tools, such as the renormalization group procedure~\cite{2011Chen}.}, gapped &Disordered, fully many-body localized\\
  &&\\ 
 Ansatz &
 \begin{tikzpicture}[scale = 0.8, baseline=(current  bounding  box.center)]


\foreach \x in {-1.4,0,1.4,2.8,4.2}{
\draw[thick] (5.3+\x,0) -- (6.7+\x,0);
\draw[thick] (6+\x,0) -- (6+\x,0.7);
\draw[thick,fill=white] (5.7+\x,-0.3) rectangle (6.3+\x,0.3);		
}

\draw[white,fill=white] (3.9,-0.7) rectangle (4.7,0.7);
\draw[white,fill=white] (10.3,-0.7) rectangle (10.9,0.7);

\coordinate[label=right:$|\tilde \psi \rangle \ {=}$] (A) at (2.4,0);

\coordinate[label=right:$A$] (A) at (4.3,0);
\coordinate[label=right:$A$] (A) at (5.65,0);
\coordinate[label=right:$A$] (A) at (7.05,0);
\coordinate[label=right:$\ldots$] (A) at (8.45,0);
\coordinate[label=right:$A$] (A) at (9.85,0);

\draw[white] (2,-2) -- (2,-2);  

\end{tikzpicture}  
 & 
\begin{tikzpicture}[scale=0.9]     
\coordinate[label=right:$\tilde U \ \  {=}$] (A) at (-2.5,0.5);
\foreach \x in {0,1.6,3.2,4.8}{
\draw[thick](-0.6+\x,-0.6) -- (-0.6+\x,1.4);
\draw[thick](-0.2+\x,-0.6) -- (-0.2+\x,1.4);
\draw[thick](0.2+\x,-0.6) -- (0.2+\x,1.4);
\draw[thick](0.6+\x,-0.6) -- (0.6+\x,1.4);
\draw[thick,fill=white] (-0.6+\x,-0.25) rectangle (0.6+\x,0.25);		
}
\foreach \x in {0.8,2.4,4}{
\draw[thick,fill=white] (-0.6+\x,0.55) rectangle (0.6+\x,1.05);		
}
\draw[white,fill=white] (-0.8,0.55) rectangle (0,1.05);	
\draw[white,fill=white] (5,0.55) rectangle (5.8,1.05);	
\draw[thick] (-0.8,0.55) -- (-0.2,0.55) -- (-0.2,1.05) -- (-0.8,1.05);
\draw[thick] (5.8,0.55) -- (5,0.55) -- (5,1.05) -- (5.8,1.05); 
\coordinate[label=right:$u_1$] (A) at (-0.35,-0.05);
\coordinate[label=right:$u_2$] (A) at (1.25,-0.05);
\coordinate[label=right:$\ldots$] (A) at (2.85,-0.05);
\coordinate[label=right:$u_{N/\ell}$] (A) at (4.35,-0.05);
\coordinate[label=right:$v_1$] (A) at (0.45,0.75);
\coordinate[label=right:$v_2$] (A) at (2.1,0.75);
\coordinate[label=right:$\ldots$] (A) at (3.7,0.75);
\coordinate[label=right:$v_{N/\ell}$] (A) at (5,0.78);
\coordinate[label=right:$v_{N/\ell}$] (A) at (-1.15,0.78);

\end{tikzpicture} \\  
 Range & Bond dimension $D$ & Length of unitary gates $\ell$  ($D = 2^{\ell/2}$)   \\
 &&\\
 Time reversal symmetry&$\mathpzc{v}^{\otimes N} |\tilde \psi^* \rangle = e^{i \theta}|\tilde \psi\rangle$&$\mathpzc{v}^{\otimes N} \tilde U^* = \tilde U \Theta$\\
 &&\\
 Local symmetry&
 \begin{tikzpicture}[scale = 1.05, baseline=(current  bounding  box.center)]

\draw[thick] (2.3,0) -- (3.7,0);
\draw[thick] (3,0) -- (3,1.3);
\draw[thick,fill=white] (3,0.8) ellipse (0.25 and 0.25);
\coordinate[label=above:$\mathpzc{v}$] (A) at (3,0.55);
\draw[thick,fill=white] (2.7,-0.3) rectangle (3.3,0.3);		
\coordinate[label=right:$A^*$] (A) at (2.7,0);

\coordinate[label=right:${=}$] (A) at (4,0);

\draw[thick] (4.6,0) -- (7.4,0);
\draw[thick] (6,0) -- (6,0.7);
\draw[thick,fill=white] (5.7,-0.3) rectangle (6.3,0.3);		

\coordinate[label=right:$A$] (A) at (5.75,0);
\draw[thick,fill=white] (5.2,0) ellipse (0.3 and 0.3);
\draw[thick,fill=white] (6.8,0) ellipse (0.3 and 0.3);
\coordinate[label=left:$w^\dg$] (A) at (5.5,0);
\coordinate[label=right:$w$] (A) at (6.6,-0.05);

\coordinate[label=right:$e^{i \phi} $] (A) at (7.5,0);

\end{tikzpicture}
 &
\begin{tikzpicture}[scale = 1.05, baseline=(current  bounding  box.center)]

\draw[thick] (1.3,0) -- (2.7,0);
\draw[ultra thick] (2,-0.7) -- (2,1.5);
\draw[thick,fill=white] (2,0.9) ellipse (0.35 and 0.35);
\coordinate[label=above:$\mathpzc{v}^{\otimes \ell}$] (A) at (2,0.65);
\draw[thick,fill=white] (1.7,-0.3) rectangle (2.3,0.3);		
\coordinate[label=right:$A^*_k$] (A) at (1.7,0);

\coordinate[label=right:${=}$] (A) at (3,0);

\draw[thick] (4,0) -- (8,0);
\draw[ultra thick] (6,-0.7) -- (6,0.7);
\draw[thick,fill=white] (5.7,-0.3) rectangle (6.3,0.3);		

\coordinate[label=right:$A_k$] (A) at (5.7,0);
\draw[thick,fill=white] (4.8,0) ellipse (0.45 and 0.45);
\draw[thick,fill=white] (7.2,0) ellipse (0.45 and 0.45);
\coordinate[label=left:$w_{2k-1}^\dg$] (A) at (5.35,0);
\coordinate[label=right:$w_{2k+1}$] (A) at (6.7,0);

\coordinate[label=right:$e^{i \phi_k} $] (A) at (8.2,0);

\end{tikzpicture}
\\
&&\\
Topological index&$w w^* = \pm \mathbb{1}$& $w_j w_j^* = \pm \mathbb{1}$ (same for all $j$)\\ 
 &&\\
Consequence&four-fold degeneracy of the ground state&four-fold degeneracy of the entanglement spectra\\ 
&entanglement spectrum for $w w^* = -\mathbb{1}$&of all eigenstates for $w_j w_j^* = -\mathbb{1}$\\
 &&\\
 \hline
\end{tabular}
\caption{Table showing the correspondence between MPS descriptions of ground states and quantum circuit descriptions of the entire set of eigenstates of MBL sytems.} \label{table}
\end{table}
\end{center}

\end{widetext}

The quantum circuits used for the proof are of the form~\cite{Wahl2017PRX}
%
%
\begin{equation}
\begin{tikzpicture}[scale=0.9]     
\coordinate[label=right:$\tilde U \ \  {=}$] (A) at (-2.5,0.5);
\foreach \x in {0,1.6,3.2,4.8}{
\draw[thick](-0.6+\x,-0.6) -- (-0.6+\x,1.4);
\draw[thick](-0.2+\x,-0.6) -- (-0.2+\x,1.4);
\draw[thick](0.2+\x,-0.6) -- (0.2+\x,1.4);
\draw[thick](0.6+\x,-0.6) -- (0.6+\x,1.4);
\draw[thick,fill=white] (-0.6+\x,-0.25) rectangle (0.6+\x,0.25);		
}
\foreach \x in {0.8,2.4,4}{
\draw[thick,fill=white] (-0.6+\x,0.55) rectangle (0.6+\x,1.05);		
}
\draw[white,fill=white] (-0.8,0.55) rectangle (0,1.05);	
\draw[white,fill=white] (5,0.55) rectangle (5.8,1.05);	
\draw[thick] (-0.8,0.55) -- (-0.2,0.55) -- (-0.2,1.05) -- (-0.8,1.05);
\draw[thick] (5.8,0.55) -- (5,0.55) -- (5,1.05) -- (5.8,1.05); 
\coordinate[label=right:$u_1$] (A) at (-0.35,-0.05);
\coordinate[label=right:$u_2$] (A) at (1.25,-0.05);
\coordinate[label=right:$\ldots$] (A) at (2.85,-0.05);
\coordinate[label=right:$u_{N/\ell}$] (A) at (4.35,-0.05);
\coordinate[label=right:$v_1$] (A) at (0.45,0.75);
\coordinate[label=right:$v_2$] (A) at (2.1,0.75);
\coordinate[label=right:$\ldots$] (A) at (3.7,0.75);
\coordinate[label=right:$v_{N/\ell}$] (A) at (5,0.78);
\coordinate[label=right:$v_{N/\ell}$] (A) at (-1.15,0.78);

\end{tikzpicture}, \label{eq:quantum_cicuit}        
\end{equation}
where $u_k$ and $v_k$ are unitaries (indicated by boxes) acting on $\ell$ sites. Each leg corresponds to a tensor index of dimension 2, i.e., in the above case $\ell = 4$. The lower dangling legs correspond to the approximate l-bit basis $l_1, l_2, \ldots, l_N$  and the upper open legs to the local physical basis. Connected legs indicate summation over the corresponding indices. $\tilde U$ approximately diagonalizes the Hamiltonian. The error of the optimized approximation decreases exponentially with $\ell$. 



For non-degenerate ground states of clean systems, time reversal symmetry implies
\begin{align}
\mc{T} |\psi\rangle = \mathpzc{v}^{\otimes N} |\psi^*\rangle = e^{i \theta} |\psi\rangle.
\end{align}
This generalizes to 
\begin{align}
\mathpzc{v}^{\otimes N} |\psi^*_{l_1 \ldots l_N}\rangle = e^{i \theta_{l_1 \ldots l_N}} |\psi_{l_1 \ldots l_N}\rangle
\end{align}
for MBL systems with eigenstates $|\psi_{l_1 \ldots l_N}\rangle$ and non-degenerate energies (possible energy degeneracies can be removed by adding infinitesimally small perturbations). Since quantum circuits with long gates form an efficient apprixmation, the same must be true for the approximate eigenstates $|\tilde \psi_{l_1 \ldots l_N} \rangle$ contained in the unitary $\tilde U$,
\begin{align}
\mathpzc{v}^{\otimes N} |\tilde \psi^*_{l_1 \ldots l_N}\rangle = e^{i \theta_{l_1 \ldots l_N}} |\tilde \psi_{l_1 \ldots l_N}\rangle.
\end{align}
For $\tilde U$ this impies
\begin{align}
\tilde U \Theta = \mathpzc{v}^{\otimes N} \tilde U^*, \label{eq:unitary_sym}
\end{align}
where $\Theta$ is the diagonal matrix with elements $e^{i \theta_{l_1 \ldots l_N}}$. 
$\Theta^{1/2}$ can be absorbed into the two-layer quantum circuit (see Sec. IV for the precise reason for this), i.e., $\tilde U \rightarrow \tilde U \Theta^{1/2}$, such that 
\begin{align}
\tilde U = \mathpzc{v}^{\otimes N} \tilde U^*. \label{eq:cool}
\end{align}
The absorption of such phase factors only works for time reversal symmetry, which is what currently precludes a generalization to on-site symmetries characterized by a symmetry group $G$.
In graphical notation, Eq.~\eqref{eq:cool} reads (we combine groups of $\ell/2$ lines into single lines with dimension $2^{\ell/2}$)
%
%
\begin{equation}
\begin{tikzpicture}[scale=0.95,baseline=(current  bounding  box.center)]    
\foreach \x in {0,1.6,3.2,4.8}{
\draw[thick](-0.4+\x,-0.6) -- (-0.4+\x,1.4);
\draw[thick](0.4+\x,-0.6) -- (0.4+\x,1.4);
\draw[thick,fill=white] (-0.4+\x,-0.25) rectangle (0.4+\x,0.25);		
}
\foreach \x in {0.8,2.4,4}{
\draw[thick,fill=white] (-0.4+\x,0.55) rectangle (0.4+\x,1.05);		
}
\draw[thick] (-0.8,0.55) -- (-0.4,0.55) -- (-0.4,1.05) -- (-0.8,1.05);
\draw[thick] (5.6,0.55) -- (5.2,0.55) -- (5.2,1.05) -- (5.6,1.05); 
\coordinate[label=right:$u_1$] (A) at (-0.3,0);
\coordinate[label=right:$u_2$] (A) at (1.3,0);
\coordinate[label=right:$\ldots$] (A) at (2.9,0);
\coordinate[label=right:$u_n$] (A) at (4.5,0);
\coordinate[label=right:$v_1$] (A) at (0.5,0.8);
\coordinate[label=right:$v_2$] (A) at (2.1,0.8);
\coordinate[label=right:$\ldots$] (A) at (3.7,0.8);
\coordinate[label=right:$v_n$] (A) at (5.15,0.78);
\coordinate[label=right:$v_n$] (A) at (-1,0.78);

\begin{scope}[shift={(0,-3.8)}]
\coordinate[label=right:${=}$] (A) at (-2,0.8);
\foreach \x in {0,1.6,3.2,4.8}{
\draw[thick](-0.4+\x,-0.6) -- (-0.4+\x,2.3);
\draw[thick](0.4+\x,-0.6) -- (0.4+\x,2.3);
\draw[thick,fill=white] (-0.4+\x,1.7) ellipse (0.3 and 0.3);
\draw[thick,fill=white] (0.4+\x,1.7) ellipse (0.3 and 0.3);
\coordinate[label=above:$\mathcal{V}$] (A) at (-0.4+\x,1.45);
\coordinate[label=above:$\mathcal{V}$] (A) at (0.4+\x,1.45);
\draw[thick,fill=white] (-0.4+\x,-0.25) rectangle (0.4+\x,0.25);		
}
\foreach \x in {0.8,2.4,4}{
\draw[thick,fill=white] (-0.4+\x,0.55) rectangle (0.4+\x,1.05);		
}
\draw[thick] (-0.8,0.55) -- (-0.4,0.55) -- (-0.4,1.05) -- (-0.8,1.05);
\draw[thick] (5.6,0.55) -- (5.2,0.55) -- (5.2,1.05) -- (5.6,1.05); 
\coordinate[label=right:$u_1^*$] (A) at (-0.3,0);
\coordinate[label=right:$u_2^*$] (A) at (1.3,0);
\coordinate[label=right:$\ldots$] (A) at (2.9,0);
\coordinate[label=right:$u_n^*$] (A) at (4.5,0);
\coordinate[label=right:$v_1^*$] (A) at (0.5,0.8);
\coordinate[label=right:$v_2^*$] (A) at (2.1,0.8);
\coordinate[label=right:$\ldots$] (A) at (3.7,0.8);
\coordinate[label=right:$v_n^*$] (A) at (5.15,0.78);
\coordinate[label=right:$v_n^*$] (A) at (-1,0.78);
\end{scope}

\end{tikzpicture}  \label{eq:qu_circuits_sym} 
\end{equation}        
with $\mathcal{V} = \mathpzc{v}^{\otimes \ell/2}$.  Note that multiplication from left to right in algebraic notation corresponds to top to bottom in graphical notation. 
If we define $u_k' = u_k^*$ and $v_k' = (\mathcal{V} \otimes \mathcal{V}) v_k^*$, we discern that Eq.~\eqref{eq:qu_circuits_sym} equates two two-layer quantum circuits,
\begin{equation}
\begin{tikzpicture}[scale=0.95,baseline=(current  bounding  box.center)]    
\foreach \x in {0,1.6,3.2,4.8}{
\draw[thick](-0.4+\x,-0.6) -- (-0.4+\x,1.4);
\draw[thick](0.4+\x,-0.6) -- (0.4+\x,1.4);
\draw[thick,fill=white] (-0.4+\x,-0.25) rectangle (0.4+\x,0.25);		
}
\foreach \x in {0.8,2.4,4}{
\draw[thick,fill=white] (-0.4+\x,0.55) rectangle (0.4+\x,1.05);		
}
\draw[thick] (-0.8,0.55) -- (-0.4,0.55) -- (-0.4,1.05) -- (-0.8,1.05);
\draw[thick] (5.6,0.55) -- (5.2,0.55) -- (5.2,1.05) -- (5.6,1.05); 
\coordinate[label=right:$u_1$] (A) at (-0.3,0);
\coordinate[label=right:$u_2$] (A) at (1.3,0);
\coordinate[label=right:$\ldots$] (A) at (2.9,0);
\coordinate[label=right:$u_n$] (A) at (4.5,0);
\coordinate[label=right:$v_1$] (A) at (0.5,0.8);
\coordinate[label=right:$v_2$] (A) at (2.1,0.8);
\coordinate[label=right:$\ldots$] (A) at (3.7,0.8);
\coordinate[label=right:$v_n$] (A) at (5.15,0.78);
\coordinate[label=right:$v_n$] (A) at (-1,0.78);

\begin{scope}[shift={(0,-2.8)}]
\coordinate[label=right:${=}$] (A) at (-2,0.8);
\foreach \x in {0,1.6,3.2,4.8}{
\draw[thick](-0.4+\x,-0.6) -- (-0.4+\x,1.4);
\draw[thick](0.4+\x,-0.6) -- (0.4+\x,1.4);
\draw[thick,fill=white] (-0.4+\x,-0.25) rectangle (0.4+\x,0.25);		
}
\foreach \x in {0.8,2.4,4}{
\draw[thick,fill=white] (-0.4+\x,0.55) rectangle (0.4+\x,1.05);		
}
\draw[thick] (-0.8,0.55) -- (-0.4,0.55) -- (-0.4,1.05) -- (-0.8,1.05);
\draw[thick] (5.6,0.55) -- (5.2,0.55) -- (5.2,1.05) -- (5.6,1.05); 
\coordinate[label=right:$u_1'$] (A) at (-0.3,0);
\coordinate[label=right:$u_2'$] (A) at (1.3,0);
\coordinate[label=right:$\ldots$] (A) at (2.9,0);
\coordinate[label=right:$u_n'$] (A) at (4.5,0);
\coordinate[label=right:$v_1'$] (A) at (0.5,0.8);
\coordinate[label=right:$v_2'$] (A) at (2.1,0.8);
\coordinate[label=right:$\ldots$] (A) at (3.7,0.8);
\coordinate[label=right:$v_n'$] (A) at (5.15,0.78);
\coordinate[label=right:$v_n'$] (A) at (-1,0.78);
\end{scope}

\end{tikzpicture}  \label{eq:qu_circuits}. 
\end{equation}        
If we multiply both sides from the bottom by $u_k'^\dg$ for $k = 1, \ldots, n$ and from the top by $v_k^\dg$, we arrive at
%
%
\begin{equation}
\begin{tikzpicture}[scale=0.9,baseline=(current  bounding  box.center)]    
\foreach \x in {0,1.6,3.2,4.8}{
\draw[thick](-0.4+\x,-0.75) -- (-0.4+\x,1.75);
\draw[thick](0.4+\x,-0.75) -- (0.4+\x,1.75);
\draw[thick,fill=white] (-0.4+\x,-0.25) rectangle (0.4+\x,0.25);		
}
 
\coordinate[label=right:${u_1'}^\dg$] (A) at (-0.3,0);
\coordinate[label=right:${u_2'}^\dg$] (A) at (1.3,0);
\coordinate[label=right:$\ldots$] (A) at (2.9,0);
\coordinate[label=right:${u_n'}^\dg$] (A) at (4.5,0);

\foreach \x in {0,1.6,3.2,4.8}{
\draw[thick,fill=white] (-0.4+\x,0.75) rectangle (0.4+\x,1.25);		
}

\coordinate[label=right:$u_1$] (A) at (-0.3,1);
\coordinate[label=right:$u_2$] (A) at (1.3,1);
\coordinate[label=right:$\ldots$] (A) at (2.9,1);
\coordinate[label=right:$u_{n}$] (A) at (4.5,1);


\begin{scope}[shift={(0,-4.2)}]
\coordinate[label=right:${=}$] (A) at (-2,1.4);

\foreach \x in {0,1.6,3.2,4.8}{
\draw[thick](-0.4+\x,0) -- (-0.4+\x,1.4);
\draw[thick](0.4+\x,0) -- (0.4+\x,1.4);
}
\foreach \x in {0.8,2.4,4}{
\draw[thick,fill=white] (-0.4+\x,0.55) rectangle (0.4+\x,1.05);		
}
\draw[thick] (-0.8,0.55) -- (-0.4,0.55) -- (-0.4,1.05) -- (-0.8,1.05);
\draw[thick] (5.6,0.55) -- (5.2,0.55) -- (5.2,1.05) -- (5.6,1.05); 
\coordinate[label=right:${v_1'}$] (A) at (0.5,0.8);
\coordinate[label=right:${v_2'}$] (A) at (2.1,0.8);
\coordinate[label=right:$\ldots$] (A) at (3.7,0.8);
\coordinate[label=right:${v_n'}$] (A) at (5.15,0.78);
\coordinate[label=right:${v_n'}$] (A) at (-1,0.78);

\begin{scope}[shift={(0,-0.6)}]
\foreach \x in {0,1.6,3.2,4.8}{
\draw[thick](-0.4+\x,2) -- (-0.4+\x,3.4);
\draw[thick](0.4+\x,2) -- (0.4+\x,3.4);
}
\foreach \x in {0.8,2.4,4}{
\draw[thick,fill=white] (-0.4+\x,2.35) rectangle (0.4+\x,2.85);		
}
\draw[thick] (-0.8,2.35) -- (-0.4,2.35) -- (-0.4,2.85) -- (-0.8,2.85);
\draw[thick] (5.6,2.35) -- (5.2,2.35) -- (5.2,2.85) -- (5.6,2.85); 

\coordinate[label=right:$v_1^\dg$] (A) at (0.5,2.61);
\coordinate[label=right:$v_2^\dg$] (A) at (2.1,2.61);
\coordinate[label=right:$\ldots$] (A) at (3.7,2.58);
\coordinate[label=right:$v_{n}^\dg$] (A) at (5.15,2.65);
\coordinate[label=right:$v_{n}^\dg$] (A) at (-1.1,2.6);
\end{scope}

\end{scope}
\end{tikzpicture} \ . \label{eq:tensor_product} 
\end{equation}        

The left hand side is a tensor product of $u_k u_k'^\dg$ and the right hand side of $v_k^\dg v_k'$ but shifted by one site with respect to each other. Hence, it has to hold that
%
%
\begin{equation}
\begin{tikzpicture}[scale=0.9,baseline=(current  bounding  box.center)]    

\draw[thick](-0.4,-1.4) -- (-0.4,1.4);
\draw[thick](0.4,-1.4) -- (0.4,1.4);
\draw[thick,fill=white] (-0.4,0.3) rectangle (0.4,0.8);		
\draw[thick,fill=white] (-0.4,-0.8) rectangle (0.4,-0.3);		

\coordinate[label=right:$u_k$] (A) at (-0.3,0.55);
\coordinate[label=right:${u_k'}^\dg$] (A) at (-0.3,-0.55);

\coordinate[label=right:${=}$] (A) at (1,0);

\begin{scope}[shift={(3.4,0)}]
\draw[thick](-0.7,-1.4) -- (-0.7,1.4);
\draw[thick](0.7,-1.4) -- (0.7,1.4);
\draw[thick,fill=white] (-0.7,0) ellipse (0.55 and 0.55);
\draw[thick,fill=white] (0.7,0) ellipse (0.55 and 0.55);
\coordinate[label=right:$w_{2k-1}$] (A) at (-1.3,0);
\coordinate[label=right:$w_{2k}$] (A) at (0.2,0);

\end{scope}

\end{tikzpicture} \label{eq:gauge1}
\end{equation}        
and
\begin{equation}
\begin{tikzpicture}[scale=0.9,baseline=(current  bounding  box.center)]    

\draw[thick](-0.4,-1.4) -- (-0.4,1.4);
\draw[thick](0.4,-1.4) -- (0.4,1.4);
\draw[thick,fill=white] (-0.4,0.3) rectangle (0.4,0.8);		
\draw[thick,fill=white] (-0.4,-0.8) rectangle (0.4,-0.3);		

\coordinate[label=right:$v_k^\dg$] (A) at (-0.3,0.55);
\coordinate[label=right:$v_k'$] (A) at (-0.3,-0.55);

\coordinate[label=right:${=}$] (A) at (1,0);

\begin{scope}[shift={(3.4,0)}]
\draw[thick](-0.7,-1.4) -- (-0.7,1.4);
\draw[thick](0.7,-1.4) -- (0.7,1.4);
\draw[thick,fill=white] (-0.7,0) ellipse (0.55 and 0.55);
\draw[thick,fill=white] (0.7,0) ellipse (0.55 and 0.55);
\coordinate[label=right:$w_{2k}$] (A) at (-1.2,0);
\coordinate[label=right:$w_{2k+1}$] (A) at (0.1,0);

\coordinate[label=right:$e^{i \phi_k}$] (A) at (1.6,0.1);

\end{scope} \label{eq:gauge2}
\end{tikzpicture},
\end{equation}        
where the $w_j$ are unitaries. The phase factor $e^{i \phi_k}$ arises because the decomposition of Eq.~\eqref{eq:tensor_product}  into a prodcut of tensors acting on blocks of $\frac{\ell}{2}$ sites is unique up to overall factors (which have to be of magnitude 1 due to unitarity). We call Eqs.~\eqref{eq:gauge1},~\eqref{eq:gauge2} a \textit{gauge transformation}, as it leaves the overall quantum circuit invariant. If we insert back the specific case of $u_k' = u_k^*$ and $v_k' = (\mathcal{V} \otimes \mathcal{V}) v_k^*$, we obtain
%
%
\begin{equation}
\begin{tikzpicture}[scale=1.1,baseline=(current  bounding  box.center)]    
\draw[thick](0.4,-0.75) -- (0.4,0.75);
\draw[thick](1.2,-0.75) -- (1.2,0.75);
\draw[thick,fill=white] (0.4,-0.25) rectangle (1.2,0.25);		
\coordinate[label=right:$u_k$] (A) at (0.5,0);

\coordinate[label=right:${=}$] (A) at (1.55,0);

\draw[thick](2.4,-0.75) -- (2.4,1.75);
\draw[thick](3.2,-0.75) -- (3.2,1.75);
\draw[thick,fill=white] (2.4,1) ellipse (0.4 and 0.4);
\draw[thick,fill=white] (3.2,1) ellipse (0.3 and 0.3);
\draw[thick,fill=white] (2.4,-0.25) rectangle (3.2,0.25);		
\coordinate[label=right:$u_k^*$] (A) at (2.5,0);
\coordinate[label=above:$w_{2k-1}$] (A) at (2.4,0.75);
\coordinate[label=above:$w_{2k}$] (A) at (3.25,0.75);

\coordinate[label=right:${,}$] (A) at (3.6,0);

\begin{scope}[shift={(4.2,0)}]

\draw[thick](0.4,-0.75) -- (0.4,1.75);
\draw[thick](1.2,-0.75) -- (1.2,1.75);
\draw[thick,fill=white] (0.4,1) ellipse (0.3 and 0.3);
\draw[thick,fill=white] (1.2,1) ellipse (0.3 and 0.3);
\coordinate[label=above:$\mathcal{V}$] (A) at (0.4,0.75);
\coordinate[label=above:$\mathcal{V}$] (A) at (1.2,0.75);
\draw[thick,fill=white] (0.4,-0.25) rectangle (1.2,0.25);		
\coordinate[label=right:$v_k^*$] (A) at (0.5,0);

\coordinate[label=right:${=}$] (A) at (1.55,0);

\draw[thick](2.4,-1.75) -- (2.4,0.75);
\draw[thick](3.2,-1.75) -- (3.2,0.75);
\draw[thick,fill=white] (2.4,-0.25) rectangle (3.2,0.25);		
\coordinate[label=right:$v_k$] (A) at (2.5,0);
\draw[thick,fill=white] (2.4,-1) ellipse (0.3 and 0.3);
\draw[thick,fill=white] (3.2,-1) ellipse (0.4 and 0.4);
\coordinate[label=below:$w_{2k}$] (A) at (2.4,-0.8);
\coordinate[label=below:$w_{2k+1}$] (A) at (3.2,-0.8);

\coordinate[label=right:$e^{i \phi_k}.$] (A) at (3.3,0);

\end{scope}

\end{tikzpicture} \label{eq:syms}
\end{equation}
If one takes the complex conjugate of Eqs.~\eqref{eq:syms} and inserts that back into the original Eqs.~\eqref{eq:syms}, one obtains
%
%
\begin{equation}
\begin{tikzpicture}[scale=1.2,baseline=(current  bounding  box.center)]    
\draw[thick](0.8,-0.75) -- (0.8,1.75);
\draw[thick](1.2,-0.75) -- (1.2,1.75);

\coordinate[label=right:${=}$] (A) at (1.4,0.5);

\draw[thick](2.4,-0.75) -- (2.4,1.75);
\draw[thick](3.4,-0.75) -- (3.4,1.75);
\draw[thick,fill=white] (2.4,1) ellipse (0.4 and 0.4);
\draw[thick,fill=white] (3.4,1) ellipse (0.4 and 0.4);
\coordinate[label=above:$w_{2k-1}$] (A) at (2.4,0.75);
\coordinate[label=above:$w_{2k}$] (A) at (3.4,0.75);
\draw[thick,fill=white] (2.4,0) ellipse (0.4 and 0.4);
\draw[thick,fill=white] (3.4,0) ellipse (0.4 and 0.4);
\coordinate[label=above:$w_{2k-1}^*$] (A) at (2.4,-0.25);
\coordinate[label=above:$w_{2k}^*$] (A) at (3.4,-0.25);

\coordinate[label=right:${,}$] (A) at (4,0.5);

\begin{scope}[shift={(4,0)}]

\draw[thick](0.8,-0.75) -- (0.8,1.75);
\draw[thick](1.2,-0.75) -- (1.2,1.75);

\coordinate[label=right:${=}$] (A) at (1.4,0.5);

\draw[thick](2.4,-0.75) -- (2.4,1.75);
\draw[thick](3.4,-0.75) -- (3.4,1.75);

\draw[thick,fill=white] (2.4,1) ellipse (0.4 and 0.4);
\draw[thick,fill=white] (3.4,1) ellipse (0.4 and 0.4);
\coordinate[label=above:$w_{2k}^*$] (A) at (2.4,0.75);
\coordinate[label=above:$w_{2k+1}^*$] (A) at (3.4,0.75);
\draw[thick,fill=white] (2.4,0) ellipse (0.4 and 0.4);
\draw[thick,fill=white] (3.4,0) ellipse (0.4 and 0.4);
\coordinate[label=above:$w_{2k}$] (A) at (2.4,-0.25);
\coordinate[label=above:$w_{2k+1}$] (A) at (3.4,-0.25);

\end{scope}

\end{tikzpicture} \label{eq:uv_sym}
\end{equation}
using $\mc V \mc V^* = \pm \mathbb{1}$. 
The left equation implies $w_{2k-1} w_{2k-1}^* = \mathbb{1} e^{i \beta_k}$, $w_{2k} w_{2k}^* = \mathbb{1} e^{-i \beta_k}$ and the right one $w_{2k} w_{2k}^* = \mathbb{1} e^{i \beta_k'}$, $w_{2k+1} w_{2k+1}^* = \mathbb{1} e^{-i \beta_k'}$. We thus have a single phase $\beta$, $w_{2k-1} w_{2k-1}^* = \mathbb{1} e^{i \beta}$, $w_{2k} w_{2k}^* = \mathbb{1} e^{-i \beta}$, for all $k = 1, \ldots, n$. Inserting the resulting $w_{2k-1} = e^{i \beta} w_{2k-1}^\top$ into itself~\cite{Pollmann2010} yields $e^{2i \beta} = 1$, i.e., $w_j w_j^* = \pm \mathbb{1}$ with the same sign for all $j = 1, 2, \ldots, 2n$. This is the topological sign of the SPT MBL phase: It does not depend on the site index $k$, i.e., it is the same for the entire chain. One cannot adiabatically change a unitary quantum circuit from a topological index $-1$ to a $+1$ index, as continuous variation of the unitaries $\{u_k, \, v_k\}$ corresponds according to Eqs.~\ref{eq:syms} to continuous variaton of $\{w_j\}$, which leaves the sign of $w_j w_j^* = \pm \mathbb{1}$ invariant. This indicates that under adiabatic perturbations of the Hamiltonian, it is impossible to connect the two phases unless the description in terms of local integrals of motion and thus in terms of quantum circuits breaks down. At such a transition point, at least one integral of motion must become delocalized.

Finally, to gain an intuition as to why one of the SPT phases has four-fold degeneracy of all eigenstates, it is illustrative to write $\tilde U$ as a matrix product operator (MPO),
%
%
\begin{equation}
\begin{tikzpicture}[scale = 0.95, baseline=(current  bounding  box.center)]


\foreach \x in {-1.4,0,1.4,2.8,4.2}{
\draw[thick] (5.3+\x,0) -- (6.7+\x,0);
\draw[ultra thick] (6+\x,-0.7) -- (6+\x,0.7);
\draw[thick,fill=white] (5.7+\x,-0.3) rectangle (6.3+\x,0.3);		
}

\draw[white,fill=white] (3.9,-0.7) rectangle (4.7,0.7);
\draw[white,fill=white] (10.3,-0.7) rectangle (10.9,0.7);

\coordinate[label=right:$\tilde U \ \ {=}$] (A) at (2.9,0);

\coordinate[label=right:$A_{n}$] (A) at (4.3,0);
\coordinate[label=right:$A_1$] (A) at (5.65,0);
\coordinate[label=right:$A_2$] (A) at (7.05,0);
\coordinate[label=right:$\ldots$] (A) at (8.45,0);
\coordinate[label=right:$A_{n}  \ {,}$] (A) at (9.85,0);

\end{tikzpicture}  \label{eq:MPO}
\end{equation}
%
%
\begin{equation}
\begin{tikzpicture}[scale=0.8,baseline=(current  bounding  box.center)]

\draw[thick] (-3.9,0.55) -- (-2.3,0.55);
\draw[ultra thick] (-3.1,1.25) -- (-3.1,-0.15);
\draw[thick,fill=white] (-3.5,0.25) rectangle (-2.7,0.85);		
\coordinate[label=right:$A_k$] (A) at (-3.5,0.55);

\coordinate[label=right:${=}$] (A) at (-1.9,0.55);
\draw[thick](-0.4,-0.55) -- (-0.4,0.55);
\draw[thick](0.4,-0.55) -- (0.4,1.65);
\draw[thick,fill=white] (-0.4,-0.25) rectangle (0.4,0.25);		
\draw[thick,fill=white] (0.4,0.85) rectangle (1.2,1.35);		

\draw[thick] (-0.8,0.55) -- (-0.4,0.55);
\draw[thick] (1.6,0.55) -- (1.2,0.55) -- (1.2,1.65);

\coordinate[label=right:$u_k$] (A) at (-0.3,0);
\coordinate[label=right:$v_k$] (A) at (0.5,1.1);

\end{tikzpicture} \ \ ,  \label{eq:A_k}
\end{equation} 
where we use thick lines to denote the combination of two vertical legs to one with dimension $2^{\ell}$. 
Eq.~\eqref{eq:syms} gives
%
%
\begin{equation}
\begin{tikzpicture}[scale=0.79,baseline=(current  bounding  box.center)]

\draw[thick](-6.4,-0.55) -- (-6.4,0.55);
\draw[thick](-5.6,-0.55) -- (-5.6,2.55);
\draw[thick,fill=white] (-6.4,-0.25) rectangle (-5.6,0.25);		
\draw[thick,fill=white] (-5.6,0.85) rectangle (-4.8,1.35);	
	
\draw[thick] (-6.8,0.55) -- (-6.4,0.55);
\draw[thick] (-4.4,0.55) -- (-4.8,0.55) -- (-4.8,2.55);
\draw[thick,fill=white] (-4.8,2) ellipse (0.3 and 0.3);
\draw[thick,fill=white] (-5.6,2) ellipse (0.3 and 0.3);
\coordinate[label=above:$\mathcal{V}$] (A) at (-4.8,1.7);
\coordinate[label=above:$\mathcal{V}$] (A) at (-5.6,1.7);

\coordinate[label=right:$u_k^*$] (A) at (-6.3,0);
\coordinate[label=right:$v_k^*$] (A) at (-5.5,1.1);

\coordinate[label=right:${=} \ \ \pm $] (A) at (-3.9,0.55);
\draw[thick](-0.4,-0.55) -- (-0.4,0.55);
\draw[thick](0.4,-0.55) -- (0.4,1.65);
\draw[thick,fill=white] (-0.4,-0.25) rectangle (0.4,0.25);		
\draw[thick,fill=white] (0.4,0.85) rectangle (1.2,1.35);

\draw[thick] (-2.4,0.55) -- (-0.4,0.55);
\draw[thick] (3.1,0.55) -- (1.2,0.55) -- (1.2,1.65);
\draw[thick,fill=white] (-1.5,0.55) ellipse (0.6 and 0.6);
\coordinate[label=right:$w_{2k-1}^*$] (A) at (-2.2,0.55);
\draw[thick,fill=white] (2.1,0.55) ellipse (0.6 and 0.6);
\coordinate[label=right:$w_{2k+1}$] (A) at (1.4,0.55);

\coordinate[label=right:$u_k$] (A) at (-0.3,0);
\coordinate[label=right:$v_k$] (A) at (0.5,1.1);

\coordinate[label=right:$e^{i \phi_k}{,}$] (A) at (3.2,0.55);

\end{tikzpicture} \label{eq:symu_k}
\end{equation}
using $w_{2k}^* w_{2k}  = \pm \mathbb{1}$, and therefore 
%
%
\begin{equation}
\begin{tikzpicture}[scale = 1.1, baseline=(current  bounding  box.center)]

\draw[thick] (1.3,0) -- (2.7,0);
\draw[ultra thick] (2,-0.7) -- (2,1.5);
\draw[thick,fill=white] (2,0.9) ellipse (0.35 and 0.35);
\coordinate[label=above:$\mathcal{V}^{\otimes 2}$] (A) at (2,0.65);
\draw[thick,fill=white] (1.7,-0.3) rectangle (2.3,0.3);		
\coordinate[label=right:$A^*_k$] (A) at (1.7,0);

\coordinate[label=right:${=}$] (A) at (3,0);

\draw[thick] (4,0) -- (8,0);
\draw[ultra thick] (6,-0.7) -- (6,0.7);
\draw[thick,fill=white] (5.7,-0.3) rectangle (6.3,0.3);		

\coordinate[label=right:$A_k$] (A) at (5.7,0);
\draw[thick,fill=white] (4.8,0) ellipse (0.45 and 0.45);
\draw[thick,fill=white] (7.2,0) ellipse (0.45 and 0.45);
\coordinate[label=left:$w_{2k-1}^\dg$] (A) at (5.35,0);
\coordinate[label=right:$w_{2k+1}$] (A) at (6.65,0);

\coordinate[label=right:$e^{i \phi_k}. $] (A) at (8.2,0);

\end{tikzpicture} \label{eq:symA}
\end{equation}
This relation is almost identical to the one obtained for MPS representing time reversal symmetric ground states~\cite{Pollmann2010}. The only differences are the lower leg corresponding to the local l-bit configuration (making it an MPO rather than an MPS) and the breaking of translational invariance reflected by the site-dependent tensors $A_k$ and virtual symmetries $w_{2k-1}$, $w_{2k+1}$. However, since $w_j w_j^* = \pm \mathbb{1}$ for all $j$, the same conclusions can be drawn as in Ref.~\onlinecite{Pollmann2010}: Consider the case of $w_j w_j^* = -\mathbb{1}$ and a specific eigenstate by fixing the l-bit configuration, i.e., the indices of the lower legs. The entanglement spectrum of that eigenstate is encoded in a reduced density matrix defined on the virtual space (horizontal legs). Due to Eq.~\eqref{eq:symA}, it has to commute with $w_{2k-1}$ and $w_{2k+1}$. For $w_j w_j^* = -\mathbb{1}$ this implies that the spectrum of the reduced density matrix has to be four-fold degenerate. Since this conclusion can be drawn independently of the chosen l-bit configuration, all eigenstates must have four-fold degenerate entanglement spectra. 

We thus showed that in the presence of time reversal symmetry, MBL systems fall into one of two topologically distinct phases, which can be distinguished by the entanglement spectra of the individual eigenstates. This is in analogy to the classification of matrix product states with time reversal symmetry~\cite{Pollmann2010,2011Chen,2011Schuch}. Along these lines, we expect a classification by the second cohomology group if the system is invariant under an on-site symmetry given by a certain symmetry group~\cite{Chandran2014SPT}. The technical problems with this extension can be gathered from the following Section. Finally, note that the derivations here only apply to bosonic systems; for fermionic systems another symmetry constraint (parity) would have to be imposed~\cite{Bultnick2017}.

The rigorous demonstration of the results above is the subject of the following Section.

%
%

\section{Theorem and Proof} 

\subsection{Theorem}

\textit{If for all sufficiently large $N$ the following conditions are fulfilled 
\begin{enumerate}
\item there exists a unitary $U$ diagonalizing the Hamiltonian $H$ defining $\tau_i^z = U \sigma_i^z U^\dg$ and a two-layer quantum circuit $\tilde U$ with $\tilde \tau_i^z = \tilde U \sigma_i^z \tilde U^\dg$ such that
$\|\tilde \tau_i^z - \tau_i^z\|_\mr{op} < c \, e^{-\frac{\ell}{\xi_i}}$ with $\xi_\mr{max} := \max_i \xi_i < c' N^{1-\mu}$ for some fixed $c, c' > 0$ and $0 < \mu < 1$ {(efficient approximability)}
\item the Hamiltonian is invariant under time reversal operation $\mc{T} = K \mathpzc{v}^{\otimes N}$, $H = \mc{T} H \mc{T}^\dg$ {(time-reversal symmetry)}
\item conditions 1 and 2 are also fulfilled for the Hamiltonian $H + \epsilon V$ with arbitrary infinitesimally small strictly local perturbations, $\epsilon \rightarrow 0$ {(MBL stability),}
\end{enumerate}
then the following holds in the thermodynamic limit ($N \rightarrow \infty$)
\begin{enumerate}
\item the Hamiltonian belongs to one of two topological classes, where one of them has a full set of eigenstates with four-fold degenerate entanglement spectra {(topological property)}
\item under adiabatic perturbations, the Hamiltonian cannot leave its topological class if the above conditions are fulfilled along the path {(topological stability)}.
\end{enumerate}
}

We will prove each of the two statements in turn.

\subsection{Proof of Statement 1}


We first prove the following: \\
\textit{Lemma 1. -- Condition 1 of the Theorem implies for $\ell(N) = \alpha N$ to leading order in $N$ that there exists a unitary $U'$ exactly diagonalizing the Hamiltonians such that $\| U' - \tilde U\|_\mr{op} < 2^{9/4} \sqrt{\frac{c N}{3}} e^{-\frac{\alpha N^\mu}{2 c'}}$.}

\noindent\textit{Proof of Lemma 1. --} 
We set $U' = U \Phi$, where $\Phi$ (to be specified below) is a diagonal matrix whose non-vanishing elements have magnitude 1. $U'$ also diagonalizes the Hamiltonian and has the same LIOMs $\tau_i^z$. 
Condition~1 hence implies for $U'$
\begin{align}
\|\sigma_i^z - \tilde U^\dg U' \sigma_i^z {U'}^\dg \tilde U\|_\mr{op} < c \, e^{-\frac{\ell}{\xi_i}}. \label{eq:sigma_blocking}
\end{align}
We write $\tilde U^\dg U'$ in blocks corresponding to degenerate subspaces of $\sigma_1^z$, 
$\tilde U^\dg U' : = \left(\begin{array}{cc}
U_{11}&U_{12}\\
U_{21}&U_{22}
\end{array}\right)$. Then, Eq.~\eqref{eq:sigma_blocking} results in
\begin{align}
&\left|\left| \left(\begin{array}{cc}
U_{11} U_{11}^\dg - U_{12} U_{12}^\dg - \mathbb{1}& U_{11} U_{21}^\dg - U_{12} U_{22}^\dg \\
U_{21} U_{11}^\dg - U_{22} U_{12}^\dg & U_{21} U_{21}^\dg - U_{22} U_{22}^\dg - \mathbb{1}
\end{array}\right) \right|\right|_\mr{op} \notag \\
&< c \, e^{-\frac{\ell}{\xi_i}}. 
\end{align}
Since $\tilde U^\dg U'$ is unitary,
\begin{align}
\left(\begin{array}{cc}
U_{11} U_{11}^\dg + U_{12} U_{12}^\dg & U_{11} U_{21}^\dg + U_{12} U_{22}^\dg \\
U_{21} U_{11}^\dg + U_{22} U_{12}^\dg & U_{21} U_{21}^\dg + U_{22} U_{22}^\dg
\end{array}\right) = \left(\begin{array}{cc}
\mathbb{1}&0 \\
0&\mathbb{1}
\end{array}\right),
\end{align}
we get
\begin{align}
&\left|\left|\left(\begin{array}{cc}
-2 U_{12} U_{12}^\dg & 2 U_{11} U_{21}^\dg \\
-2 U_{22} U_{12}^\dg & 2 U_{21} U_{21}^\dg 
\end{array}\right) \right|\right|_\mr{op} \notag \\
&= 2 \left|\left| \left(\begin{array}{cc}
U_{12} & U_{11} \\
U_{22} & U_{21} 
\end{array}\right)
\left(\begin{array}{cc}
-U_{12}^\dg&0\\
0&U_{21}^\dg
\end{array}\right)
\right|\right|_\mr{op} \notag \\
&= 2 \max\left(\| U_{12}\|_\mr{op}, \| U_{21}\|_\mr{op}\right)
 < c \, e^{-\frac{\ell}{\xi_i}}. \label{eq:off-diagonal}
\end{align}
We define $D$ as the diagonal matrix with the same diagonal elements as $\tilde U^\dg U'$. 
By definition, the operator norm of $D - \tilde U^\dg U'$ (with $|\cdot | = \| \cdot \|_2$) is
\begin{align}
&\| D - \tilde U^\dg U' \|_\mr{op}^2 \notag \\
&= \max_{|v_1|^2 + |v_2|^2 = 1} \left| \left(\begin{array}{cc}
D_{11} - U_{11}& -U_{12} \\
U_{21} & D_{22} - U_{22} 
\end{array}\right) \left(\begin{array}{c} \mb v_1 \\ \mb v_2 \end{array}\right) \right|^2
\notag  \\
&\leq \left( |(D_{11} - U_{11}) \mb v_1| + |U_{12} \mb v_2|\right)^2 + \left( |U_{21} \mb v_1| + |(D_{22} - U_{22}) \mb v_2|\right)^2 \notag \\
&\leq \left(\|D_{11} - U_{11} \|_\mr{op} |\mb v_1 | + \| U_{12}\|_\mr{op} \sqrt{1 - |\mb v_1|^2}\right)^2 \notag \\
&+ \left( \| U_{21} \|_\mr{op} |\mb v_1| + \|D_{22} - U_{22}\|_\mr{op} \sqrt{1 - |\mb v_1|^2}\right)^2 \notag \\
&= \left(\|D_{11} - U_{11} \|_\mr{op}^2 + \| U_{21} \|_\mr{op}^2\right) |\mb v_1|^2 + 2( \|D_{11} - U_{11} \|_\mr{op} \| U_{12}\|_\mr{op} \notag \\
&+  \| U_{21} \|_\mr{op} \|D_{22} - U_{22}\|_\mr{op} ) |\mb v_1 | \sqrt{1-|\mb v_1|^2} \notag \\
&+ \left(\| U_{12}\|_\mr{op}^2 + \|D_{22} - U_{22}\|_\mr{op}^2\right) (1 - |\mb v_1|^2). \label{eq:bound1}
\end{align}
The second term of the last expression is maximal at $|\mb v_1| = \frac{1}{\sqrt{3}}$, the other ones are maximal at $|\mb v_1| = 0,1$, respectively, which allows us to bound
\begin{align}
&\| D - \tilde U^\dg U' \|_\mr{op}^2 \notag \\
&\leq \frac{2 \sqrt{2}}{3}( \|D_{11} - U_{11} \|_\mr{op} \| U_{12}\|_\mr{op} +  \| U_{21} \|_\mr{op} \|D_{22} - U_{22}\|_\mr{op} ) \notag \\
&+ \max\left(\|D_{11} - U_{11} \|_\mr{op}^2 + \| U_{21} \|_\mr{op}^2,\| U_{12}\|_\mr{op}^2 + \|D_{22} - U_{22}\|_\mr{op}^2\right). \label{eq:bound2}
\end{align}
Using Eq.~\eqref{eq:off-diagonal} and $\|D_{11} - U_{11}\|_\mr{op}, \|D_{22} - U_{22}\|_\mr{op} < 2$, this implies
\begin{align}
\| \mathbb{1} - \tilde U^\dg U' \|_\mr{op}^2 &< \frac{4 \sqrt{2} c }{3} e^{-\frac{\ell}{\xi_1}} + \frac{c^2}{4} e^{-\frac{2 \ell}{\xi_1}} \notag \\
&+ \max\left(\|D_{11} - U_{11}\|^2_\mr{op}, \|D_{22} - U_{22}\|^2_\mr{op}\right).
\end{align}
Assume now $\| D_{11} - U_{11} \|_\mr{op} \geq \|D_{22} - U_{22} \|_\mr{op}$ (the treatment of the opposite case is analogous). If we decompose $U_{11}$ into blocks according to the invariant subspaces of $\sigma_2^z$,
\begin{align}
U_{11} = \left(\begin{array}{cc}
U_{11,11}& U_{11,12} \\
U_{11,21}& U_{11,22} 
\end{array}\right),
\end{align}
Eq.~\eqref{eq:off-diagonal} for a $2 \times 2$ block form of $\tilde U^\dg U'$ with respect to $\sigma_2^z$ implies $\|U_{11,12}\|_\mr{op}, \| U_{11,21}\|_\mr{op} < \frac{c}{2} e^{-\frac{\ell}{\xi_2}}$. This has to hold, because the operator norm of a matrix block cannot be larger than the one of the overall matrix. We can thus repeat the bounding approach of Eqs.~\eqref{eq:bound1}, \eqref{eq:bound2} for $\|D_{11} - U_{11}\|^2_\mr{op}$ and obtain
\begin{align}
&\|D - \tilde U^\dg U' \|_\mr{op}^2 < \frac{4 \sqrt{2} c }{3} \left( e^{-\frac{\ell}{\xi_1}} + e^{-\frac{\ell}{\xi_2}}\right) + \frac{c^2}{4} \left(e^{-\frac{2 \ell}{\xi_1}} + e^{-\frac{2 \ell}{\xi_2}} \right) \notag \\
&+ \max\left(\|D_{11,11} - U_{11,11}\|^2_\mr{op}, \|D_{11,22} - U_{11,22}\|^2_\mr{op}\right).
\end{align}
Continuation of the same procedure for the remaining sites $i = 3, 4, \ldots, N$ yields
\begin{align}
\|D - \tilde U^\dg U' \|_\mr{op}^2 < \frac{4 \sqrt{2} c }{3} \sum_{i=1}^N e^{-\frac{\ell}{\xi_i}} + \frac{c^2}{4} \sum_{i=1}^N e^{-\frac{2 \ell}{\xi_i}} := \gamma. \label{eq:final_bound}
\end{align}
This implies $\| [D]_{:n} - [\tilde U^\dg U']_{:n}| < \sqrt{\gamma}$, where $[\ldots]_{:n}$ refers to the $n$-th column vector. This requires
\begin{align}
1 - \sqrt{\gamma} < |[D]_{:n}| = |D_{nn}| < 1 + \sqrt{\gamma}
\end{align} 
We choose $\Phi$ such that the diagonal elements $D_{nn}$ of $\tilde U^\dg U' = \tilde U^\dg U \Phi$ are non-negative. 
Therefore, we obtain using the triangluar inequality
\begin{align}
&\|\mathbb{1} - \tilde U^\dg U' \|_\mr{op} \leq \|\mathbb{1} - D\|_\mr{op} + \|D - \tilde U^\dg U' \|_\mr{op} \notag \\
&< 2 \sqrt{\gamma} = 2 \sqrt{\frac{4 \sqrt{2} c }{3} \sum_{i=1}^N e^{-\frac{\ell}{\xi_i}} + \frac{c^2}{4} \sum_{i=1}^N e^{-\frac{2 \ell}{\xi_i}}}.
\end{align}
Thus, in the limit $\ell/\xi_\mr{max} \rightarrow \infty$
\begin{align}
\|\mathbb{1} - \tilde U^\dg U' \|_\mr{op} < 4 \sqrt{\frac{\sqrt{2} c}{3} N} e^{-\frac{\ell}{2 \xi_\mr{max}}}.
\end{align}
For $\ell = \alpha N$ and $\xi_\mr{max} < c' N^{1-\mu}$, we arrive at
\begin{align}
\|U' - \tilde U \|_\mr{op} < 2^{9/4} \sqrt{\frac{c N}{3}} e^{-\frac{\alpha N^\mu}{2 c'}}. \label{eq:U-U}
\end{align}
\qed


We set $\delta(N) := 2^{9/4} \sqrt{\frac{c N}{3}} e^{-\frac{\alpha N^\mu}{2 c'}}$ with obviously $\lim_{N \rightarrow \infty} \delta(N) = 0$. Next, we derive the implications of time reversal symmetry on the unitary matrix $U'$. (For simplicity, we drop the prime from now on.) By its definition,
\begin{align}
U^\dg H U = E, \label{eq:eigs}
\end{align}
where $E$ is the diagonal matrix containing the corresponding energy eigenvalues. Condition 2 says $H = \mathpzc{v}^{\otimes N} H^* ({\mathpzc{v}^\dg})^{\otimes N}$, i.e., we have 
\begin{align}
U^\dg \mathcal{V} H^* \mathcal{V}^\dg U = E,
\end{align}
where in the following we use the symbol $\mathcal{V}$ in a slightly sloppy way denoting the right number of tensor products of unitaries $\mathpzc{v}$ (i.e., here $\mathcal{V} = \mathpzc{v}^{\otimes N}$). 
Due to $H^* = U^* E U^\top$ we obtain 
\begin{align}
U^\dg \mathcal{V} U^* E U^\top \mathcal{V}^\dg U = E. \label{eq:diag-diag}
\end{align}
For spin systems, time reversal symmetry does not protect any degeneracies. Condition 3 allows us to add a local perturbation that breaks any remaining accidental degeneracies for finite $N$, such as $V =  \sum_i h_i' \sigma_z^i$ with random $h_i'$ for Hamiltonian Eq.~\eqref{eq:top_Ham}. Hence, we can assume that there are no degeneracies, such that Eq.~\eqref{eq:diag-diag} implies
\begin{align}
\Theta = U^\dg \mathcal{V} U^*,
\end{align}
where $\Theta$ is a diagonal matrix with diagonal elements of magnitude 1. Using Lemma 1, the triangular inequality yields
\begin{align}
\|\tilde U \Theta - \mathcal{V} \tilde U^* \|_\mr{op} &\leq \|U \Theta - \mathcal{V} U^*\|_\mr{op} + \|(\tilde U - U ) \Theta\|_\mr{op} \notag \\
&+ \| \mathcal{V} ( U^* - \tilde U^*)\|_\mr{op} \notag \\
&< 0 + \delta(N) \|\Theta\|_\mr{op} + \delta(N) \|\mathcal{V}\|_\mr{op} \notag \\
&= 2 \delta(N). \label{eq:bound-symmetry}
\end{align}

\input{Lemma2}

From the preceding discussion we gather that $\Theta$ cannot be absorbed into the definition of $\tilde U$ in the case of an on-site symmetry (instead of time reversal symmetry), which prevents extension of the current proof to all one-dimensional symmetry protected MBL phases. 

Let us now define $\delta_1(N) = (2 + 4N) \delta(N)$ (which also vanishes in the limit $N \rightarrow \infty$).
In the new quantum circuit, each unitary $\overline u_k$ has $3 \ell/2$ (upper and lower) left legs and $5 \ell/2$ right legs and vice versa for the $\overline v_k$'s. 

\input{Lemma3}

The entanglement spectrum is given by the ``entanglement energies'', which are the eigenvalues of the entanglement Hamiltonian $H_\mr{ent}$ defined by $\rho_L = e^{-H_\mr{ent}}$. $\rho_L$ is the reduced density matrix obtained after tracing out half of the chain from a certain eigenstate.

\input{Lemma4}

\subsection{Proof of Statement 2}

We assume $N$ to be finite, and strict topological protection will again follow in the limit $N \rightarrow \infty$. Suppose the adiabatic perturbation is described by a parameter $\lambda \in [0,1]$ with corresponding Hamiltonian $H(\lambda)$ such that Conditions 1 to 3 of the Theorem are fulfilled for all $\lambda$. Condition 1 thus requires the existence of a unitary $U(\lambda)$ diagonalizing the Hamiltonian such that $\tau_i^z(\lambda) = U(\lambda) \sigma_i^z U^\dg(\lambda)$ fulfills $\| \tilde \tau_i^z (\lambda) - \tau_i^z(\lambda) \|_\mr{op} < c(\lambda) e^{-\frac{\ell}{\xi_i(\lambda)}}$ with $\tilde \tau_i^z(\lambda) = \tilde U (\lambda) \sigma_i^z \tilde U^\dg (\lambda) = \overline U (\lambda) \sigma_i^z \overline U^\dg(\lambda)$. Due to Condition 3, we can assume that $H(\lambda)$ (at least after some infintely small $\lambda$-independent perturbation $\epsilon V$) is non-degenerate for almost all $\lambda$. Degeneracies only appear at level crossings, which are isolated points for finite $N$. 

First of all, note that one can always define a unitary $U_\mr{cont}(\lambda)$ which diagonalizes the overall Hamiltonian $H(\lambda)$ and changes continuously as a function of $\lambda$. This can be seen by comparing the two limits $\lim_{\epsilon \rightarrow 0 \pm} H(\lambda + \epsilon)$ expressed in terms of $U_\mr{cont}(\lambda +\epsilon)$ and $E(\lambda + \epsilon)$. For almost all $\lambda$, $U_\mr{cont}(\lambda)$ has to be related to $U(\lambda)$ via
\begin{align}
U(\lambda) = U_\mr{cont}(\lambda) P(\lambda)
\end{align}
where $P(\lambda)$ is a permutation matrix whose non-vanishing elements have arbitrary phases (and magnitude 1). 
According to Lemmas 1 and 2 we thus have
\begin{align}
\| U_\mr{cont}(\lambda) P(\lambda) - \overline U(\lambda)\|_\mr{op} < \delta_1(N). \label{eq:cont}
\end{align}
Now consider two points $\lambda_1, \lambda_2 \in [0,1]$. We want to show that the topological index of the corresponding quantum circuits $\overline U(\lambda_1)$, $\overline U(\lambda_2)$ is the same. Due to the triangular inequality,
\begin{align}
&\|\overline U(\lambda_1) - \overline U(\lambda_2) P^\dg(\lambda_2) P(\lambda_1) \|_\mr{op} \notag \\ 
&\leq \|\overline U(\lambda_1) - U_\mr{cont}(\lambda_1) P(\lambda_1) \|_\mr{op} \notag \\
&+ \|U_\mr{cont}(\lambda_1) P(\lambda_1) - U_\mr{cont}(\lambda_2) P(\lambda_1) \|_\mr{op} \notag \\
&+ \| U_\mr{cont}(\lambda_2) P(\lambda_1) - \overline U(\lambda_2) P^\dg(\lambda_2) P(\lambda_1) \|_\mr{op} \notag \\
&< 2 \delta_1(N) + \|U_\mr{cont}(\lambda_1) - U_\mr{cont}(\lambda_2)\|_\mr{op}
\end{align}
due to Eq.~\eqref{eq:cont}. We choose $\lambda_2 := \lambda_1 + \epsilon$, such that because of the continuity of $U_\mr{cont}(\lambda)$
\begin{align}
\|\overline U(\lambda_1) - \overline U(\lambda_2) P^\dg(\lambda_2) P(\lambda_1) \|_\mr{op} < 2 \delta_1(N) + \mc{O}(\epsilon), \label{eq:bound_perm}
\end{align}
where the $\mc{O}(\epsilon)$ term is $N$-independent. We consider the quantum circuit defined by $\overline U_{21} := \overline U^\dg(\lambda_2) \overline U(\lambda_1)$, which is due to Eq.~\eqref{eq:bound_perm} close to a permutation matrix. That quantum circuit reads graphically (we denote the unitaries of $\overline U(\lambda_1)$ by $\overline u_k$, $\overline v_k$ and those of $\overline U(\lambda_2)$ by $\hat u_k$, $\hat v_k$
%
%
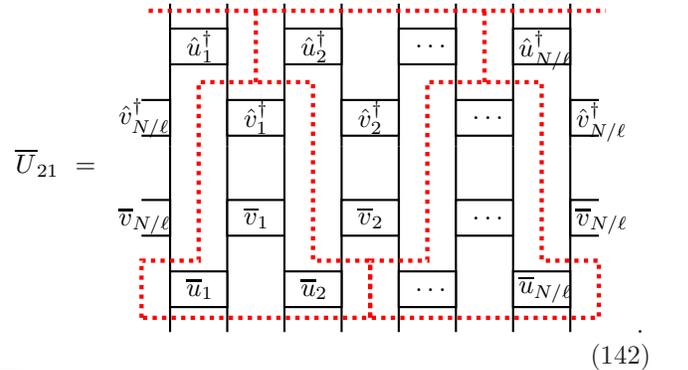
\begin{equation}
\begin{tikzpicture}[scale=0.95]     

\coordinate[label=above:${\overline U_{21} \ =}$] (A) at (-2,1.45);

\foreach \x in {0,1.6,3.2,4.8}{
\draw[thick](-0.4+\x,-0.6) -- (-0.4+\x,2);
\draw[thick](0.4+\x,-0.6) -- (0.4+\x,2);
\draw[thick,fill=white] (-0.4+\x,-0.25) rectangle (0.4+\x,0.25);		
}
\foreach \x in {0.8,2.4,4}{
\draw[thick,fill=white] (-0.4+\x,0.75) rectangle (0.4+\x,1.25);		
}
\draw[thick] (-0.8,0.75) -- (-0.4,0.75) -- (-0.4,1.25) -- (-0.8,1.25);
\draw[thick] (5.6,0.75) -- (5.2,0.75) -- (5.2,1.25) -- (5.6,1.25); 
\coordinate[label=right:$\overline u_1$] (A) at (-0.3,0);
\coordinate[label=right:$\overline u_2$] (A) at (1.3,0);
\coordinate[label=right:$\ldots$] (A) at (2.9,0);
\coordinate[label=right:$\overline u_{N/\ell}$] (A) at (4.35,0);
\coordinate[label=right:$\overline v_1$] (A) at (0.5,1);
\coordinate[label=right:$\overline v_2$] (A) at (2.1,1);
\coordinate[label=right:$\ldots$] (A) at (3.7,1);
\coordinate[label=right:$\overline v_{N/\ell}$] (A) at (5.15,0.95);
\coordinate[label=right:$\overline v_{N/\ell}$] (A) at (-1.25,0.95);

\foreach \x in {0,1.6,3.2,4.8}{
\draw[thick](-0.4+\x,2) -- (-0.4+\x,4);
\draw[thick](0.4+\x,2) -- (0.4+\x,4);
\draw[thick,fill=white] (-0.4+\x,3.15) rectangle (0.4+\x,3.65);		
}
\foreach \x in {0.8,2.4,4}{
\draw[thick,fill=white] (-0.4+\x,2.15) rectangle (0.4+\x,2.65);		
}
\draw[thick] (-0.8,2.15) -- (-0.4,2.15) -- (-0.4,2.65) -- (-0.8,2.65);
\draw[thick] (5.6,2.15) -- (5.2,2.15) -- (5.2,2.65) -- (5.6,2.65); 
\coordinate[label=right:${\hat u_1^\dg}$] (A) at (-0.3,3.4);
\coordinate[label=right:${\hat u_2^\dg}$] (A) at (1.3,3.4);
\coordinate[label=right:$\ldots$] (A) at (2.9,3.4);
\coordinate[label=right:${\hat u_{N/\ell}^\dg}$] (A) at (4.35,3.35);
\coordinate[label=right:${\hat v_1^\dg}$] (A) at (0.5,2.4);
\coordinate[label=right:${\hat v_2^\dg}$] (A) at (2.1,2.4);
\coordinate[label=right:$\ldots$] (A) at (3.7,2.4);
\coordinate[label=right:${\hat v_{N/\ell}^\dg}$] (A) at (5.15,2.35);
\coordinate[label=right:${\hat v_{N/\ell}^\dg}$] (A) at (-1.25,2.4);

\draw[ultra thick,dotted,red] (-0.8,-0.4) -- (-0.8,0.4) -- (0,0.4) -- (0,2.9) -- (1.6,2.9) -- (1.6,0.4) -- (2.4,0.4) -- (2.4,-0.4) -- cycle;
\draw[ultra thick,dotted,red] (2.4,-0.4) -- (2.4,0.4) -- (3.2,0.4) -- (3.2,2.9) -- (4.8,2.9) -- (4.8,0.4) -- (5.6,0.4) -- (5.6,-0.4) -- cycle;
\draw[ultra thick,dotted,red] (-0.7,3.9) -- (5.7,3.9); 
\draw[ultra thick,dotted,red] (0.8,2.9) -- (0.8,3.9);
\draw[ultra thick,dotted,red] (4,2.9) -- (4,3.9);

\end{tikzpicture} .
\end{equation}
$\overline U_{21}$ can also be written as a two-layer quantum circuit if the unitaries are blocked together as indicated by red dashed lines. From Lemma 2, we know that $\| \overline U(\lambda_1) - \mc{V} \overline U^*(\lambda_1) \|_\mr{op} < \delta_1(N)$ and $\| \overline U(\lambda_2) - \mc{V} \overline U^*(\lambda_2) \|_\mr{op} < \delta_1(N)$, which together with Eq.~\eqref{eq:ADE} implies
\begin{align}
\| \overline U_{21} - \overline U_{21}^* \|_\mr{op} < 2 \delta_1(N).
\end{align}
Hence, we can apply Lemma 3, which in this case states that

\noindent
%
%
\begin{tikzpicture}[scale=1.25]     
\coordinate[label=right:$\|$] (A) at (-0.9,1);
\foreach \x in {0,1.6}{
\draw[thick](-0.4+\x,-0.6) -- (-0.4+\x,3.3);
\draw[thick](0.4+\x,-0.6) -- (0.4+\x,3.3);
\draw[thick,fill=white] (-0.4+\x,-0.25) rectangle (0.4+\x,0.25);		
}
\draw[thick,fill=white] (0.4,0.55) rectangle (1.2,1.05);		
\draw[thick,fill=white] (0.4,1.5) rectangle (1.2,2);		

\coordinate[label=right:$\overline u_{2m-1}^*$] (A) at (-0.45,0);
\coordinate[label=right:$\overline u_{2m}^*$] (A) at (1.25,0);
\coordinate[label=right:$\overline v_{2m-1}^*$] (A) at (0.35,0.8);
\coordinate[label=right:$\hat v_{2m-1}^\top$] (A) at (0.35,1.75);

\coordinate[label=right:${-}$] (A) at (2.3,1);

\begin{scope}[shift={(3.4,0)}]

\foreach \x in {0,1.6}{
\draw[thick](-0.4+\x,-0.6) -- (-0.4+\x,3.3);
\draw[thick](0.4+\x,-0.6) -- (0.4+\x,3.3);
\draw[thick,fill=white] (-0.4+\x,-0.25) rectangle (0.4+\x,0.25);		
}
\draw[thick,fill=white] (0.4,0.55) rectangle (1.2,1.05);		
\draw[thick,fill=white] (0.4,1.5) rectangle (1.2,2);		

\draw[thick,fill=white] (0,2.7) ellipse (0.5 and 0.5); 
\draw[thick,fill=white] (1.6,2.7) ellipse (0.5 and 0.5); 
\coordinate[label=above:$\tilde W_{2m-1}^*$] (A) at (0,2.45);
\coordinate[label=above:$\tilde W_{2m}^*$] (A) at (1.6,2.45);

\coordinate[label=right:$\overline u_{2m-1}$] (A) at (-0.45,0);
\coordinate[label=right:$\overline u_{2m}$] (A) at (1.25,0);
\coordinate[label=right:$\overline v_{2m-1}$] (A) at (0.35,0.8);
\coordinate[label=right:$\hat v_{2m-1}^\dg$] (A) at (0.35,1.75);

\coordinate[label=right:$\|_\mr{op}$] (A) at (2.1,1);

\end{scope}

\end{tikzpicture}  
\begin{align}
< 2 \delta_1(N) \label{eq:W}
\end{align}
with $\| \mathbb{1} \mp \tilde W_k \tilde W_k^* \|_\mr{op} < 22 \delta_1(N)$. On the other hand, Lemma 3 applied to the individual quantum circuits $\overline U(\lambda_1)$, $\overline U(\lambda_2)$ implies to leading order using Eq.~\eqref{eq:ADE}
($\hat w_j$ for $\overline U(\lambda_2)$ corresponds to $\tilde w_j$ for $\overline U(\lambda_1)$)

\noindent
%
%
\begin{tikzpicture}[scale=1.35]     
\coordinate[label=right:$\|$] (A) at (-0.8,1);
\foreach \x in {0,1.6}{
\draw[thick](-0.4+\x,-0.6) -- (-0.4+\x,3.3);
\draw[thick](0.4+\x,-0.6) -- (0.4+\x,3.3);
\draw[thick,fill=white] (-0.4+\x,-0.25) rectangle (0.4+\x,0.25);		
}
\draw[thick,fill=white] (0.4,0.55) rectangle (1.2,1.05);		
\draw[thick,fill=white] (0.4,1.5) rectangle (1.2,2);		

\coordinate[label=right:$\overline u_{2m-1}^*$] (A) at (-0.45,0);
\coordinate[label=right:$\overline u_{2m}^*$] (A) at (1.25,0);
\coordinate[label=right:$\overline v_{2m-1}^*$] (A) at (0.35,0.8);
\coordinate[label=right:$\hat v_{2m-1}^\top$] (A) at (0.35,1.75);

\coordinate[label=right:${-}$] (A) at (2.1,1);

\begin{scope}[shift={(3,0)}]

\foreach \x in {0,1.6}{
\draw[thick](-0.4+\x,-0.6) -- (-0.4+\x,3.3);
\draw[thick](0.4+\x,-0.6) -- (0.4+\x,3.3);
\draw[thick,fill=white] (-0.4+\x,-0.25) rectangle (0.4+\x,0.25);		
}
\draw[thick,fill=white] (0.4,0.55) rectangle (1.2,1.05);		
\draw[thick,fill=white] (0.4,1.5) rectangle (1.2,2);		

\draw[thick,fill=white] (-0.4,2.7) ellipse (0.38 and 0.38); 
\draw[thick,fill=white] (0.4,2.7) ellipse (0.38 and 0.38);
\draw[thick,fill=white] (1.2,2.7) ellipse (0.38 and 0.38);  
\draw[thick,fill=white] (2,2.7) ellipse (0.38 and 0.38); 
\coordinate[label=above:$\tilde w_{4m-3}^*$] (A) at (-0.4,2.45);
\coordinate[label=above:$\hat w_{4m-2}^\dg$] (A) at (0.4,2.45);
\coordinate[label=above:$\hat w_{4m-1}^\dg$] (A) at (1.2,2.45);
\coordinate[label=above:$\tilde w_{4m}^*$] (A) at (2,2.45);

\coordinate[label=right:$\overline u_{2m-1}$] (A) at (-0.45,0);
\coordinate[label=right:$\overline u_{2m}$] (A) at (1.25,0);
\coordinate[label=right:$\overline v_{2m-1}$] (A) at (0.35,0.8);
\coordinate[label=right:$\hat v_{2m-1}^\dg$] (A) at (0.35,1.75);

\coordinate[label=right:$\times$] (A) at (2,1);
\end{scope}

\coordinate[label=right:${\times \frac{\gamma_{2m-1}^*}{\hat \gamma_{2m-1}^*} \|_\mr{op}}$] (A) at (-0.8,-1.2);

\begin{scope}[shift={(0,-5.3)}]

\coordinate[label=right:$< \|$] (A) at (-1.1,1);
\foreach \x in {0,1.6}{
\draw[thick](-0.4+\x,-0.6) -- (-0.4+\x,3.3);
\draw[thick](0.4+\x,-0.6) -- (0.4+\x,3.3);
\draw[thick,fill=white] (-0.4+\x,-0.25) rectangle (0.4+\x,0.25);		
}
\draw[thick,fill=white] (0.4,0.55) rectangle (1.2,1.05);		
\draw[thick,fill=white] (0.4,1.5) rectangle (1.2,2);		

\coordinate[label=right:$\overline u_{2m-1}^*$] (A) at (-0.45,0);
\coordinate[label=right:$\overline u_{2m}^*$] (A) at (1.25,0);
\coordinate[label=right:$\overline v_{2m-1}^*$] (A) at (0.35,0.8);
\coordinate[label=right:$\hat v_{2m-1}^\top$] (A) at (0.35,1.75);

\coordinate[label=right:${-}$] (A) at (2.1,1);

\begin{scope}[shift={(3,0)}]

\foreach \x in {0,1.6}{
\draw[thick](-0.4+\x,-0.6) -- (-0.4+\x,3.3);
\draw[thick](0.4+\x,-0.6) -- (0.4+\x,3.3);
\draw[thick,fill=white] (-0.4+\x,-0.25) rectangle (0.4+\x,0.25);		
}
\draw[thick,fill=white] (0.4,0.55) rectangle (1.2,1.05);		
\draw[thick,fill=white] (0.4,2.45) rectangle (1.2,2.95);		

\draw[thick,fill=white] (-0.4,2.7) ellipse (0.38 and 0.38); 
\draw[thick,fill=white] (0.4,1.75) ellipse (0.3 and 0.3);
\draw[thick,fill=white] (1.2,1.75) ellipse (0.3 and 0.3);  
\draw[thick,fill=white] (2,2.7) ellipse (0.38 and 0.38); 
\coordinate[label=above:$\tilde w_{4m-3}^*$] (A) at (-0.4,2.45);
\coordinate[label=above:$\mc V^\dg$] (A) at (0.4,1.55);
\coordinate[label=above:$\mc V^\dg$] (A) at (1.2,1.55);
\coordinate[label=above:$\tilde w_{4m}^*$] (A) at (2,2.45);

\coordinate[label=right:$\overline u_{2m-1}$] (A) at (-0.45,0);
\coordinate[label=right:$\overline u_{2m}$] (A) at (1.25,0);
\coordinate[label=right:$\overline v_{2m-1}$] (A) at (0.35,0.8);
\coordinate[label=right:$\hat v_{2m-1}^\top$] (A) at (0.35,2.75);

\coordinate[label=right:$\times$] (A) at (2,1);
\end{scope}


\end{scope}

\end{tikzpicture}  
\begin{align}
\times \gamma_{2m-1}^*\|_\mr{op} + \, 2 \delta_2(N) < 4 \delta_1(N) + 4 \delta_2(N), \label{eq:w}
\end{align}
where we employed $\mc V^\dg = \pm \mc V^*$, Eqs.~\eqref{eq:Lemma3u} and~\eqref{eq:Lemma3v} (each one twice) and $|\gamma_k| \rightarrow 1$ for $N \rightarrow \infty$. 
Using the triangular inequality and Eqs.~\eqref{eq:W} and~\eqref{eq:w}, we thus arrive at
\begin{align}
&\| \tilde W_{2m-1}^* \otimes \tilde W_{2m}^*  \notag \\
 &-\frac{\gamma_{2m-1}^*}{\hat \gamma_{2m-1}^*} (\tilde w_{4m-3}^* \otimes \hat w_{4m-2}^\dg) \otimes (\hat w_{4m-1}^\dg \otimes \tilde w_{4m}^*)\|_\mr{op} \notag \\
 &< 6 \delta_1(N) + 4 \delta_2(N).
\end{align}
Hence,
\begin{align}
\| \tilde W_{2m}^* - \kappa_{2m} \hat w_{4m-1}^\dg \otimes \tilde w_{4m}^* \|_\mr{op} < 6 \delta_1(N) + 4 \delta_2(N)
\end{align}
with $\kappa_{2m}  \in \mathbb{C}$ and $|\kappa_{2m}| \rightarrow 1$ for $N \rightarrow \infty$. Finally, via Eq.~\eqref{eq:ADE}
\begin{align}
&\|\tilde W_{2m} \tilde W_{2m}^* - |\kappa_{2m}|^2 (\hat w_{4m-1}^\top \otimes \tilde w_{4m}) (\hat w_{4m-1}^\dg \otimes \tilde w_{4m}^*)\|_\mr{op} \notag \\
&= \|\tilde W_{2m} \tilde W_{2m}^* - |\kappa_{2m}|^2 (\hat w_{4m-1}^\top  \hat w_{4m-1}^\dg) \otimes (\tilde w_{4m} \tilde w_{4m}^*) \|_\mr{op}\notag \\
&< 12 \delta_1(N) + 8 \delta_2(N). \label{eq:top_index_relation}
\end{align}
Owing to Eq.~\eqref{eq:bound_perm}, 
\begin{align}
\|[\overline U_{21}]_{:l} -[ P^\dg(\lambda_2) P(\lambda_1)]_{:l}\|_2 < 2\delta_1(N) + \mc{O}(\epsilon),
\end{align}
where $[\ldots]_{:l}$ refers to the $l$-th column vector. In the limit $\epsilon \rightarrow 0$, the column vectors of $\overline U_{21}$ thus converge to the ones of $P^\dg(\lambda_2) P(\lambda_1)$ with increasing $N$, which are product states. Hence, 
 the topological index of $\overline U_{21}$ can only be $+1$, since otherwise one could use Lemma 4 to show that the column vectors of $\overline U_{21}$ have approximately two-fold degenerate entanglement spectra, which is not the case for product states. We thus have
\begin{align}
\|\mathbb{1} - \tilde W_k \tilde W_k^* \|_\mr{op} < 22 \delta_1(N),
\end{align}
i.e., according to Eqs.~\eqref{eq:top_index_relation} and~\eqref{eq:ADE}  
\begin{align}
&\| \mathbb{1} -  |\kappa_{2m}|^2 (\hat w_{4m-1}^\top \hat w_{4m-1}^\dg) \otimes (\tilde w_{4m} \tilde w_{4m}^*) \|_\mr{op} \notag \\
&< 34 \delta_1(N) + 8 \delta_2(N).
\end{align}
According to Lemma 3, $\|\mathbb{1} \mp \tilde w_j \tilde w_j^* \| < 11 \delta_1(N)$ and $\|\mathbb{1} \mp \hat w_j \hat w_j^* \| < 11 \delta_1(N)$. Therefore, for sufficiently large $N$, the topological indices of $\overline U(\lambda_1)$ and $\overline  U(\lambda_2)$ have to be identical. In the thermodynamic limit, the topological index is thus conserved along the path. \qed

As a corollary, we obtain that it is impossible to adiabatically move the system from the topologically non-trivial to the topologically trivial phase without breaking time reversal symmetry or violating the FMBL condition.

\section{Conclusions}  
We proved the existence of an SPT phase of all eigenstates of FMBL systems invariant under time reversal symmetry. Using a two-layer quantum circuit, we demonstrated the four-fold degeneracy of the entanglement spectra of all eigenstates in the SPT phase. The obtained classification thus resembles the one of one-dimensional gapped ground states with time reversal symmetry~\cite{Pollmann2010,2011Chen}. An extension to on-site symmetries and fermionic systems is expected to yield similar results, but cannot be carried out with the tools introduced in this paper only, which crucially require the absorption of the phase matrix $\Theta$ into the tensor network. 

We proved the robustness of the two phases to arbitrary symmetry preserving perturbations so long as the system remains FMBL. Thus, our results imply that there can be no symmetry-preserving transition between the topological and the trivial FMBL phase without the delocalization of at least some eigenstate(s), i.e., the system has to enter a critical regime. This behavior resembles SPT ground states in one dimension, which retain their topological properties unless the perturbations break the symmetry or close the energy gap leading to a non-local change of the ground state wave function. Note that this by no means implies that the above transition has similar features as the MBL-to-thermal transition~\cite{kjall2014many}. In particular, there might be a crossover regime, where as a function of energy some eigenstates are localized and topologically trivial, whereas others are localized and SPT, with delocalized eigenstates separating those energy windows~\cite{2015Slagle}.

Similar approaches might be employed to fully classify symmetry protected FMBL phases in one dimension. This would involve showing that there are no topological subclasses compared to the ones obtained for one-dimensional ground states, which was also not carried out in the current analysis. Note on the other hand that MBL systems cannot have non-Abelian symmetries~\cite{2016Potter_Vasseur}. 

Furthermore, the defined topological index might be used in numerical simulations with quantum circuits in order to map out the phase diagrams of MBL systems with time reversal symmetry. 
The advantage over exact diagonalization would be that the tensor network approach does not require prior knowledge of an order  parameter~\cite{kjall2014many,bahri2015localization} or the splitting between ideally degenerate energy levels for open boundary conditions to be smaller than the mean level spacing~\cite{Huse2013LPQO}. 

In two dimensions, even strongly disordered systems are believed to eventually equilibrate~\cite{chandran2016higherD,deRoeck2017Stability,ponte2017thermalinclusions}, though possibly on astronomically long time scales~\cite{ponte2017thermalinclusions}. On short time scales, they behave many-body localized~\cite{Choi1547,bordia2017quasiperiodic2D}, and are thus well described by shallow two-dimensional quantum circuits~\cite{2DMBL}. A similar approach might therefore be used to prove short-time symmetry and localization protection in two dimensions. 

\section*{Acknowledgments} The author would like to thank Steven Simon for discussions and providing valuable feedback to an earlier manuscript. The author is also grateful to Christoph S\"{u}nderhauf, Norbert Schuch, Arijeet Pal, Amos Chan, Andrea De Luca and David P\'{e}rez-Garc\'{i}a for helpful discussions. This work was supported by TOPNES, EPSRC grant number EP/I031014/1. This project has received funding from the European Union’s Horizon 2020 research and innovation programme under the Marie Skłodowska-Curie grant agreement No. 749150. The contents of this article reflect only the author's views and not the views of the European Commission. Statement of compliance with EPSRC policy framework on research data: This publication is theoretical work that does not require supporting research data.

\appendix

\bibliography{biblioMBL}{}

\section{Cluster state with random couplings} \label{sec:cluster_state}

If we set $\sigma_h = \sigma_V = 0$ in the Hamiltonian~(1), i.e.,
\begin{align}
H = \sum_{i=1}^N \lambda_i \sigma_x^{i-1} \sigma_z^i \sigma_x^{i+1}, \label{eq:Ham}
\end{align}
it is a sum of commuting local projectors (stabilizer code). As a result, the unitary matrix $U_\mr{cl}$ which diagonalizes the Hamiltonian can be written exactly as a two-layer quantum circuit~\cite{Skrovseth2009PRA} with $\ell = 2$ (assuming $N$ to be even) and
\begin{align}
u_k = v_k  = u_\mr{cl} = \frac{1}{2} \left(\begin{array}{cccc}
1&-1&-1&-1\\
-1&1&-1&-1\\
-1&-1&1&-1\\
-1&-1&-1&1
\end{array}\right).
\end{align}
In the $|\pm\rangle = \tfrac{1}{\sqrt{2}} (|0\rangle \pm |1\rangle)$ basis, $u_\mr{cl}$ is diagonal, i.e., all $u_k$ and $v_k$ extended by identities to the full Hilbert space of $N$ spins commute with each other. Note also that if we set $v_{N/2} = \mathbb{1}$, the obtained quantum circuit diagonalizes the Hamiltonian with open boundary conditions, i.e., where the sum in Eq.~\eqref{eq:Ham} extends only from 2 to $N-1$.

One can easily verify that $u_\mr{cl}$ fulfills (setting $X := \sigma_x$, $Z:= \sigma_z$, $I := \mathbb{1}_{2 \times 2}$)
\begin{align}
u_\mr{cl} &= - (I \otimes Z) u_\mr{cl} (X \otimes Z)  \label{eq:sym1} \\
                   &= - (Z \otimes X) u_\mr{cl} (Z \otimes I) \label{eq:sym2} \\
                   &= - (Z \otimes I) u_\mr{cl} (Z \otimes X) \label{eq:sym3} \\
                   &= -(X \otimes Z) u_\mr{cl} (I \otimes Z) \label{eq:sym4} \\
                   &= -(Z \otimes Z) u_\mr{cl} ((XZ) \otimes (XZ)) \label{eq:sym5} \\
                   &= -((XZ) \otimes (XZ)) u_\mr{cl} (Z \otimes Z) \label{eq:sym6},
\end{align}
where Eq.~\eqref{eq:sym5} follows from Eqs.~\eqref{eq:sym1} and~\eqref{eq:sym3} and~Eq.~\eqref{eq:sym6} from Eqs.~\eqref{eq:sym2} and~\eqref{eq:sym4}. The symmetries also reflect the fact that the Hamiltonian commutes with $Z_\mr{even} := (I \otimes Z)^{\otimes \frac{N}{2}}$ and $Z_\mr{odd} := (Z \otimes I)^{\otimes \frac{N}{2}}$.
Eqs.~\eqref{eq:sym3} and~\eqref{eq:sym4} yield
%
%
\begin{equation}
\begin{tikzpicture}[scale=1,baseline=(current  bounding  box.center)]

\draw[thick](-6.4,-0.55) -- (-6.4,0.55);
\draw[thick](-5.6,-0.55) -- (-5.6,0.55);
\draw[thick,fill=white] (-6.4,-0.25) rectangle (-5.6,0.25);		
\draw[thick,fill=white] (-5.6,0.55) rectangle (-4.8,1.05);	
	
\draw[thick] (-6.8,0.55) -- (-6.4,0.55);
\draw[thick] (-4.4,0.55) -- (-4.8,0.55);
\draw[thick] (-4.8,1.4) -- (-4.8,1.05) -- (-5.6,1.05) -- (-5.6,1.4);
\coordinate[label=above:$Z$] (A) at (-5.6,1.4);

\coordinate[label=right:$u_k$] (A) at (-6.3,0);
\coordinate[label=right:$v_k$] (A) at (-5.5,0.8);

\coordinate[label=right:${=} \ \  $] (A) at (-3,0.6);
\draw[thick](-0.4,-0.55) -- (-0.4,0.55);
\draw[thick](0.4,-0.55) -- (0.4,0.55);
\draw[thick,fill=white] (-0.4,-0.25) rectangle (0.4,0.25);		
\draw[thick,fill=white] (0.4,0.55) rectangle (1.2,1.05);		

\coordinate[label=left:$X$] (A) at (-0.8,0.55);
\draw[thick] (-0.8,0.55) -- (-0.4,0.55);
\draw[thick] (1.6,0.55) -- (1.2,0.55);
\draw[thick] (1.2,1.4) -- (1.2,1.05) -- (0.4,1.05) -- (0.4,1.4);
\coordinate[label=right:$X$] (A) at (1.6,0.55);
\coordinate[label=below:$Z$] (A) at (0.4,-0.55);

\coordinate[label=right:$u_k$] (A) at (-0.3,0);
\coordinate[label=right:$v_k$] (A) at (0.5,0.8);

\end{tikzpicture}. \label{eq:Zeven}
\end{equation}
For periodic boundary conditions this implies $Z_\mr{even} |\psi_{l_1 l_2 \ldots l_N}^{\mr{PBC}}\rangle = (-1)^{l_2 + l_4 + \ldots + l_N} |\psi_{l_1 l_2 \ldots l_N}^{\mr{PBC}}\rangle$, where $|\psi_{l_1 l_2 \ldots l_N}^{\mr{PBC}}\rangle$ is the column vector corresponding to the l-bit configuration $l_1, l_2, \ldots, l_N$ of $U_\mr{cl}^{\mr{PBC}}$ ($l_i = 0, 1$). Similarly, Eqs.~\eqref{eq:sym1} and~\eqref{eq:sym2} imply
%
%
\begin{equation}
\begin{tikzpicture}[scale=1,baseline=(current  bounding  box.center)]

\draw[thick](-6.4,-0.55) -- (-6.4,0.55);
\draw[thick](-5.6,-0.55) -- (-5.6,0.55);
\draw[thick,fill=white] (-6.4,-0.25) rectangle (-5.6,0.25);		
\draw[thick,fill=white] (-5.6,0.55) rectangle (-4.8,1.05);	
	
\draw[thick] (-6.8,0.55) -- (-6.4,0.55);
\draw[thick] (-4.4,0.55) -- (-4.8,0.55);
\draw[thick] (-4.8,1.4) -- (-4.8,1.05) -- (-5.6,1.05) -- (-5.6,1.4);
\coordinate[label=above:$Z$] (A) at (-4.8,1.4);

\coordinate[label=right:$u_k$] (A) at (-6.3,0);
\coordinate[label=right:$v_k$] (A) at (-5.5,0.8);

\coordinate[label=right:${=} \ \  $] (A) at (-3,0.6);
\draw[thick](-0.4,-0.55) -- (-0.4,0.55);
\draw[thick](0.4,-0.55) -- (0.4,0.55);
\draw[thick,fill=white] (-0.4,-0.25) rectangle (0.4,0.25);		
\draw[thick,fill=white] (0.4,0.55) rectangle (1.2,1.05);		

\coordinate[label=left:$Z$] (A) at (-0.8,0.55);
\draw[thick] (-0.8,0.55) -- (-0.4,0.55);
\draw[thick] (1.6,0.55) -- (1.2,0.55);
\draw[thick] (1.2,1.4) -- (1.2,1.05) -- (0.4,1.05) -- (0.4,1.4);
\coordinate[label=right:$Z$] (A) at (1.6,0.55);
\coordinate[label=below:$Z$] (A) at (-0.4,-0.55);

\coordinate[label=right:$u_k$] (A) at (-0.3,0);
\coordinate[label=right:$v_k$] (A) at (0.5,0.8);

\end{tikzpicture}. \label{eq:Zodd}
\end{equation}
Therefore, we have $Z_\mr{odd} |\psi_{l_1 l_2 \ldots l_N}^{\mr{PBC}}\rangle = (-1)^{l_1 + l_3 + \ldots + l_{N-1}} |\psi_{l_1 l_2 \ldots l_N}^{\mr{PBC}}\rangle$ and as a result also $\sigma_z^{\otimes N} |\psi_{l_1 l_2 \ldots l_N}^{\mr{PBC}}\rangle$ \\ $= (-1)^{\sum_{i=1}^N l_i} |\psi_{l_1 l_2 \ldots l_N}^{\mr{PBC}}\rangle$. This can be seen directly from the relation
%
%
\begin{equation}
\begin{tikzpicture}[scale=1,baseline=(current  bounding  box.center)]

\draw[thick](-6.4,-0.55) -- (-6.4,0.55);
\draw[thick](-5.6,-0.55) -- (-5.6,0.55);
\draw[thick,fill=white] (-6.4,-0.25) rectangle (-5.6,0.25);		
\draw[thick,fill=white] (-5.6,0.55) rectangle (-4.8,1.05);	
	
\draw[thick] (-6.8,0.55) -- (-6.4,0.55);
\draw[thick] (-4.4,0.55) -- (-4.8,0.55);
\draw[thick] (-4.8,1.4) -- (-4.8,1.05) -- (-5.6,1.05) -- (-5.6,1.4);
\coordinate[label=above:$Z$] (A) at (-4.8,1.4);
\coordinate[label=above:$Z$] (A) at (-5.6,1.4);

\coordinate[label=right:$u_k$] (A) at (-6.3,0);
\coordinate[label=right:$v_k$] (A) at (-5.5,0.8);

\coordinate[label=right:${=} \ \ \ \ \   $] (A) at (-3.2,0.6);
\draw[thick](-0.4,-0.55) -- (-0.4,0.55);
\draw[thick](0.4,-0.55) -- (0.4,0.55);
\draw[thick,fill=white] (-0.4,-0.25) rectangle (0.4,0.25);		
\draw[thick,fill=white] (0.4,0.55) rectangle (1.2,1.05);		

\coordinate[label=left:$ZX$] (A) at (-0.8,0.55);
\draw[thick] (-0.8,0.55) -- (-0.4,0.55);
\draw[thick] (1.2,1.4) -- (1.2,1.05) -- (0.4,1.05) -- (0.4,1.4);
\draw[thick] (1.6,0.55) -- (1.2,0.55);
\coordinate[label=right:$XZ$] (A) at (1.6,0.55);
\coordinate[label=below:$Z$] (A) at (-0.4,-0.55);
\coordinate[label=below:$Z$] (A) at (0.4,-0.55);

\coordinate[label=right:$u_k$] (A) at (-0.3,0);
\coordinate[label=right:$v_k$] (A) at (0.5,0.8);

\end{tikzpicture}. \label{eq:Z}
\end{equation}
which is obtained using Eqs.~\eqref{eq:sym5} and~\eqref{eq:sym6}. This corresponds to the virtual symmetry $w_{2k-1} = ZX$ and thus $w_j w_j^* = - \mathbb{1}$. Note that strictly speaking Eq.~\eqref{eq:Z} is not the same as Eq.~(13) in the main part, since here we have a $Z \otimes Z$ operator acting on the l-bit basis (i.e., $\Theta = \sigma_z^{\otimes N}$). These additional signs can be removed by defining the quantum circuit in terms of the unitaries $\overline u_k = u_\mr{cl} \left[\left(\begin{smallmatrix} 1 & 0 \\ 0 & i\end{smallmatrix}\right) \otimes \left(\begin{smallmatrix} 1 & 0 \\ 0 & i\end{smallmatrix}\right)\right]$ and $\overline v_k = u_\mr{cl}$. This comes  at the price of making $\overline u_k$  complex (such that the complex conjugate appears on the left hand side of Eq.~\eqref{eq:Z} in accordance with Eq.~(13)). The corresponding overall unitary matrix $\overline U_\mr{cl}$ still diagonalizes the Hamiltonian and we still have $\overline w_j \overline w_j^* = - \mathbb{1}$. 

As an add-on, let us remark that Eqs.~\eqref{eq:sym1} to~\eqref{eq:sym6} also indicate how degenerate eigenstates for open boundary conditions are related by l-bit flips at the edges. The degeneracy arises because $|\psi_{l_1 \ldots l_N}^{\mr{OBC}}\rangle$,  $Z_\mr{even} |\psi_{l_1 \ldots l_N}^{\mr{OBC}}\rangle$, $Z_\mr{odd} |\psi_{l_1 \ldots l_N}^{\mr{OBC}}\rangle$ and $\sigma_z^{\otimes N} |\psi_{l_1 \ldots l_N}^{\mr{OBC}}\rangle$ have the same energy (since $Z_\mr{even}$ and $Z_\mr{odd}$ commute with the Hamiltonian). As we will see, those states are linearly independent, i.e., the energy spectrum is four-fold degenerate. Noting that $v_{N/2} = \mathbb{1}$, we analyze the symmetries
%
%
 \begin{equation}
\begin{tikzpicture}[scale=1,baseline=(current  bounding  box.center)]

\draw[thick](-6.4,-0.55) -- (-6.4,1.4);
\draw[thick](-5.6,-0.55) -- (-5.6,0.55);
\draw[thick,fill=white] (-6.4,-0.25) rectangle (-5.6,0.25);		
\draw[thick,fill=white] (-5.6,0.55) rectangle (-4.8,1.05);	
	
\draw[thick] (-4.4,0.55) -- (-4.8,0.55);
\draw[thick] (-4.8,1.4) -- (-4.8,1.05) -- (-5.6,1.05) -- (-5.6,1.4);
\coordinate[label=above:$Z$] (A) at (-5.6,1.4);

\coordinate[label=right:$u_1$] (A) at (-6.3,0);
\coordinate[label=right:$v_1$] (A) at (-5.5,0.8);

\coordinate[label=right:${=} \ \ \ \ $] (A) at (-2.8,0.6);
\draw[thick](-0.4,-0.55) -- (-0.4,1.4);
\draw[thick](0.4,-0.55) -- (0.4,0.55);
\draw[thick,fill=white] (-0.4,-0.25) rectangle (0.4,0.25);		
\draw[thick,fill=white] (0.4,0.55) rectangle (1.2,1.05);		

\draw[thick] (1.6,0.55) -- (1.2,0.55);
\draw[thick] (1.2,1.4) -- (1.2,1.05) -- (0.4,1.05) -- (0.4,1.4);
\coordinate[label=right:$X$] (A) at (1.6,0.55);
\coordinate[label=below:$X$] (A) at (-0.4,-0.55);
\coordinate[label=below:$Z$] (A) at (0.4,-0.55);

\coordinate[label=right:$u_1$] (A) at (-0.3,0);
\coordinate[label=right:$v_1$] (A) at (0.5,0.8);

\end{tikzpicture}
\end{equation}
(obtained by using Eqs.~\eqref{eq:sym3} and~\eqref{eq:sym1}) and
%
%
\begin{equation}
\begin{tikzpicture}[scale=1,baseline=(current  bounding  box.center)]

\draw[thick](-6.4,-0.55) -- (-6.4,0.55);
\draw[thick](-5.6,-0.55) -- (-5.6,1.4);
\draw[thick,fill=white] (-6.4,-0.25) rectangle (-5.6,0.25);		
	
\draw[thick] (-6.8,0.55) -- (-6.4,0.55);
\coordinate[label=above:$Z$] (A) at (-5.6,1.4);

\coordinate[label=right:$u_{N/2}$] (A) at (-6.45,0);

\coordinate[label=right:${=} \ \ \ \ \ \ \ \ -  $] (A) at (-3.5,0.6);
\draw[thick](-0.4,-0.55) -- (-0.4,0.55);
\draw[thick](0.4,-0.55) -- (0.4,1.4);
\draw[thick,fill=white] (-0.4,-0.25) rectangle (0.4,0.25);		

\coordinate[label=left:$X$] (A) at (-0.8,0.55);
\draw[thick] (-0.8,0.55) -- (-0.4,0.55);
\coordinate[label=below:$Z$] (A) at (0.4,-0.55);

\coordinate[label=right:$u_{N/2}$] (A) at (-0.45,0);

\end{tikzpicture} \  \ ,
\end{equation}
which follows from Eq.~\eqref{eq:sym4}. Using these relations along with Eq.~\eqref{eq:Zeven}, we find $Z_\mr{even} |\psi_{l_1 l_2 \ldots l_N}^{\mr{OBC}}\rangle$ \\ $= (-1)^{1+l_2 + l_4 + \ldots + l_{N}} |\psi_{\overline l_1 l_2 \ldots l_N}^{\mr{OBC}}\rangle$ with $\overline l_i : = 1 - l_i$. Similarly, we obtain
%
%
 \begin{equation}
\begin{tikzpicture}[scale=1,baseline=(current  bounding  box.center)]

\draw[thick](-6.4,-0.55) -- (-6.4,1.4);
\draw[thick](-5.6,-0.55) -- (-5.6,0.55);
\draw[thick,fill=white] (-6.4,-0.25) rectangle (-5.6,0.25);		
\draw[thick,fill=white] (-5.6,0.55) rectangle (-4.8,1.05);	
	
\draw[thick] (-4.4,0.55) -- (-4.8,0.55);
\draw[thick] (-4.8,1.4) -- (-4.8,1.05) -- (-5.6,1.05) -- (-5.6,1.4);
\coordinate[label=above:$Z$] (A) at (-6.4,1.4);
\coordinate[label=above:$Z$] (A) at (-4.8,1.4);

\coordinate[label=right:$u_1$] (A) at (-6.3,0);
\coordinate[label=right:$v_1$] (A) at (-5.5,0.8);

\coordinate[label=right:${=} \ \ \ \ $] (A) at (-2.8,0.6);
\draw[thick](-0.4,-0.55) -- (-0.4,1.4);
\draw[thick](0.4,-0.55) -- (0.4,0.55);
\draw[thick,fill=white] (-0.4,-0.25) rectangle (0.4,0.25);		
\draw[thick,fill=white] (0.4,0.55) rectangle (1.2,1.05);		

\draw[thick] (1.6,0.55) -- (1.2,0.55);
\draw[thick] (1.2,1.4) -- (1.2,1.05) -- (0.4,1.05) -- (0.4,1.4);
\coordinate[label=right:$Z$] (A) at (1.6,0.55);
\coordinate[label=below:$Z$] (A) at (-0.4,-0.55);

\coordinate[label=right:$u_1$] (A) at (-0.3,0);
\coordinate[label=right:$v_1$] (A) at (0.5,0.8);

\end{tikzpicture}
\end{equation}
from Eqs.~\eqref{eq:sym1} and~\eqref{eq:sym2} and
%
%
\begin{equation}
\begin{tikzpicture}[scale=1,baseline=(current  bounding  box.center)]

\draw[thick](-6.4,-0.55) -- (-6.4,0.55);
\draw[thick](-5.6,-0.55) -- (-5.6,1.4);
\draw[thick,fill=white] (-6.4,-0.25) rectangle (-5.6,0.25);		
	
\draw[thick] (-6.8,0.55) -- (-6.4,0.55);

\coordinate[label=right:$u_{N/2}$] (A) at (-6.45,0);

\coordinate[label=right:${=} \ \ \ \ \ \ \ \ -  $] (A) at (-3.5,0.6);
\draw[thick](-0.4,-0.55) -- (-0.4,0.55);
\draw[thick](0.4,-0.55) -- (0.4,1.4);
\draw[thick,fill=white] (-0.4,-0.25) rectangle (0.4,0.25);		

\coordinate[label=left:$Z$] (A) at (-0.8,0.55);
\draw[thick] (-0.8,0.55) -- (-0.4,0.55);
\coordinate[label=below:$Z$] (A) at (-0.4,-0.55);
\coordinate[label=below:$X$] (A) at (0.4,-0.55);

\coordinate[label=right:$u_{N/2}$] (A) at (-0.45,0);

\end{tikzpicture} 
\end{equation}
from Eq.~\eqref{eq:sym3}. The last two equations combined with Eq.~\eqref{eq:Zodd} tell us that $Z_\mr{odd} |\psi_{l_1  \ldots l_{N-1} l_N}^{\mr{OBC}}\rangle$ \\
 $= (-1)^{1+l_1 + l_3 + \ldots + l_{N-1}} |\psi_{l_1  \ldots l_{N-1} \overline l_N}^{\mr{OBC}}\rangle$. As a result, we have $\sigma_z^{\otimes N} |\psi_{l_1 l_2 \ldots l_{N-1} l_N}^{\mr{OBC}}\rangle = (-1)^{1+\sum_{i=1}^N l_i} |\psi_{\overline l_1 l_2 \ldots l_{N-1} \overline l_N}^{\mr{OBC}}\rangle$.

\end{document}